\documentclass[
    twocolumn,
    preprintnumbers, 
    amssymb,
    amsmath,
    aps,
    prd,
    floatfix,
    nofootinbib,
    superscriptaddress,
]{revtex4-2}

\usepackage{graphicx}
\usepackage{subfigure}
\usepackage[
    colorlinks,
    linkcolor=blue,
    anchorcolor=black,
    citecolor=blue
]{hyperref}

\usepackage{multirow}

\newcommand\bperp{\ensuremath{b_\perp}}

\newcommand\kaperp{\ensuremath{ k_{1\perp}}}
\newcommand\kbperp{\ensuremath{ {k^{\prime}}_{1\perp}}}
\newcommand\kcperp{\ensuremath{ k_{2\perp}}}
\newcommand\kdperp{\ensuremath{ {k^{\prime}}_{2\perp}}}

\newcommand{\nocontentsline}[3]{}
\let\origcontentsline\addcontentsline
\newcommand\stoptoc{\let\addcontentsline\nocontentsline}
\newcommand\resumetoc{\let\addcontentsline\origcontentsline}

\begin{document}

\title{Lighting up the Photon Wigner Distribution via Dilepton Productions}

\author{Yu Shi} 
\email{yu.shi@polytechnique.edu} 
\affiliation{Key Laboratory of Particle Physics and Particle Irradiation (MOE), Institute of frontier and interdisciplinary science, Shandong University, Qingdao, Shandong 266237, China}
\affiliation{CPHT, CNRS, \'Ecole polytechnique,  Institut Polytechnique de Paris, 91120 Palaiseau, France}

\author{Lin Chen}  
\email{lin.chen@usc.es}
\affiliation{School of Science and Engineering, The Chinese University of Hong Kong, Shenzhen, Guangdong, 518172, P.R. China}
\affiliation{University of Science and Technology of China, Hefei, Anhui, 230026, P.R.China}
\affiliation{Instituto Galego de F\'isica de Altas Enerx\'ias IGFAE, Universidade de Santiago de Compostela, E-15782 Galicia-Spain}

\author{Shu-Yi Wei}  
\email{shuyi@sdu.edu.cn}
\affiliation{Key Laboratory of Particle Physics and Particle Irradiation (MOE), Institute of frontier and interdisciplinary science, Shandong University, Qingdao, Shandong 266237, China}

\author{Bo-Wen Xiao}  
\email{xiaobowen@cuhk.edu.cn}
\affiliation{School of Science and Engineering, The Chinese University of Hong Kong, Shenzhen, Guangdong, 518172, P.R. China}
\affiliation{Southern Center for Nuclear-Science Theory (SCNT), Institute of Modern Physics, Chinese Academy of Sciences, Huizhou 516000, Guangdong Province, China}

\begin{abstract}
We present a systematic investigation of lepton pair production through photon-photon fusion processes in heavy-ion collisions. It is demonstrated that the dilepton production at a given impact parameter ($b_\perp$) with a fixed transverse momentum imbalance ($q_\perp$) can be factorized into a unified formula in terms of the Wigner photon distribution of heavy nuclei. We show that this framework provides a comprehensive description of all the relevant data from RHIC to the LHC, with a strong evidence that the quasi-real photon can be radiated not only from the nucleus as a whole, standing for the coherent contribution, but also from the sub-structures inside the nucleus, representing the incoherent contribution. Further predictions are made for the anisotropies in the correlations between $q_\perp$, $b_\perp$, and the dilepton transverse momentum ($P_\perp$). This will help us to constrain the photon Wigner distribution which plays a crucial role to study the gluonic matter of nucleus at small-$x$ through the diffractive photoproduction processes in heavy ion collision.
\end{abstract}



\maketitle

\stoptoc

\section{Introduction}
\label{sec:intro}
Dilepton production in photon-induced events measured in ultra-peripheral collisions (UPCs) has become a rich physics program at the Relativistic Heavy Ion Collider (RHIC) and the Large Hadron Collider (LHC). Due to a large photon flux surrounding the relativistic heavy nuclei, one can measure the dilepton production at an abundant rate and provide detailed information for this process at various transverse momentum ($P_\perp$), the transverse momentum imbalance ($q_\perp$), and the impact parameter ($b_\perp$). In recent years, abundant high-precision experimental data on dilepton production through photon-photon fusion process ($\gamma \gamma \rightarrow l^+ l^-$) has been collected at RHIC~\cite{STAR:2004bzo, STAR:2018ldd, STAR:2019wlg, STAR:2023nos, STAR:2023vvb} and the LHC~\cite{ATLAS:2016vdy, ATLAS:2018pfw, ATLAS:2020epq, CMS:2020skx, ALICE:2022hvk, ATLAS:2022yad, ATLAS:2022srr, ATLAS:2022ryk, CMS:2022arf}. A large number of diverse observables from the experimental measurement have been provided to understand and study its production mechanism. As it provides a clean environment to study the photon distributions from heavy nuclei, this process has attracted much attention which leads to a series of theoretical researches~\cite{Brodsky:1971ud, Terazawa:1973tb, Budnev:1975poe, Bertulani:1987tz,Cahn:1990jk, Bottcher:1991nd,  Vidovic:1992ik, Hencken:1994my, Krauss:1997vr, Guclu:1999fx, Baur:2001jj, Baltz:2002pp, Baur:2003ar,Hencken:2004td, Bertulani:2005ru, Baur:2007zz, Baltz:2007kq,Baltz:2009jk, Staig:2010by, Klein:2018cjh, Klein:2018fmp, Klusek-Gawenda:2018zfz,Vysotskii:2018eic, Zha:2018tlq, Zha:2018tlq,Azevedo:2019fyz, Li:2019yzy, Li:2019sin,  Klein:2020jom,Klein:2020fmr,Xiao:2020ddm, Karadag:2019gvc, Klusek-Gawenda:2020eja,  Brandenburg:2020ozx,Hatta:2021jcd, Zha:2021jhf, Wang:2021kxm, Wang:2022ihj, Wang:2022gkd, Brandenburg:2022tna, Lin:2022flv, Shao:2022stc, Luo:2023syp, Shao:2023zge,Dai:2024imb,Zhang:2024mql} and Monte-Carlo studies~\cite{Klein:2016yzr, Harland-Lang:2018iur, Harland-Lang:2020veo, Burmasov:2021phy, Shao:2022cly, Harland-Lang:2023ohq}. Additionally, this process improves our understanding of photo-nuclear interaction and the structure of the nucleus, and is important to explore new physics beyond the standard model~\cite{DELPHI:2003nah, ATLAS:2022ryk, Knapen:2016moh, CMS:2018erd, Ellis:2017edi, Xu:2022qme, Shao:2023bga}  and study QED properties under extreme conditions~\cite{Baltz:1997di, Baur:1998ay, Klein:2020fmr, Steinberg:2021lfm,Hattori:2020htm, Copinger:2020nyx, Brandenburg:2021lnj}.
 
 {Theoretical studies employing different approaches can effectively describe various experimental observations.  However, there is still some tension between theoretical results and experimental data at the large $\alpha$/high $q_\perp$ region in the near-miss events ($b_\perp \sim 2R_A$). } Therefore, currently, different high-precision physical observables from RHIC and the LHC present a challenge to theorists: 
 to develop a unified factorized framework incorporating both impact parameter dependence and transverse momentum dependence, which can account for all experimental observables, such as dilepton anisotropies and $q_T$ broadening.  Based on previous studies\cite{Li:2019yzy, Li:2019sin, Klein:2020jom, Klein:2020fmr,Xiao:2020ddm}, we derive a generalized factorization framework on dilepton production in terms of the photon fusion hard part and the photon Wigner distributions, which provide five-dimensional photon distribution within the nuclei. We will show that this framework not only reproduces the results from previous studies\cite{Li:2019yzy, Li:2019sin, Klein:2020jom,Xiao:2020ddm}, but also reveal a range of new anisotropic correlations among the vectors $\vec{P}_\perp$, $\vec{q}_\perp$, and $\vec{b}_\perp$. These anisotropies are useful for probing the properties of linearly polarized photons in heavy nuclei.

With a detailed analysis, we also find that the quasi-real photons not only can be emitted coherently from the nucleus as a whole but also incoherently from the substructures within the nucleus. Specifically, the incoherent contribution is one of the production mechanisms responsible for $ q_{\perp} $ broadening, and it predominantly contributes to events with neutron emissions in the large-$q_\perp$ region. Thus, we provide a thorough description of existing experimental measurements from RHIC and LHC within this unified framework.

More recently, diffractive photo-production processes in heavy ion collisions, such as heavy quarkonium production, have been shown to provide a unique probe of gluon tomography in nuclei at small-$x$~\cite{Klein:2019qfb,Mantysaari:2020axf,Mantysaari:2017slo,Morreale:2021pnn,Mantysaari:2023prg,Mantysaari:2023xcu,Kovchegov:2023bvy,Cepila:2023dxn,ALICE:2021gpt,CMS:2023snh,LHCb:2022ahs,ALICE:2023jgu,STAR:2023nos}. Because these processes also depend on the incoming photon Wigner distribution, the formalism derived in our paper will help solidify the theoretical foundation for these studies. In other words, a comparative study between dilepton production through photon-photon fusion and diffractive photo-production will assist in building a rigorous framework to explore gluonic matter in cold nuclei during heavy ion collisions.

\section{The generalized factorization}
\label{sec:fact}
Let us describe the generalized factorization for the dilepton production in the two-photon fusion process as follows
\begin{equation}
\gamma(k_1)+\gamma(k_2) \rightarrow l^+(p_{1}) + l^-(p_{2}).
\end{equation}
In heavy-ion collisions, two high-energy nuclei, separated by a transverse distance $b_\perp$, emit a significant number of quasi-real photons with low virtuality. Two photons from different sources collide to produce a lepton pair with transverse momentum $p_{1\perp}$ and $p_{2\perp}$. The collision point is distanced at $b_{1\perp}$ and $b_{2\perp}$ from the two heavy nuclei, respectively. In both RHIC and the LHC experiments, measurements have shown that the transverse momentum of the final lepton is often much larger than the total momentum of the lepton pair, $\vec q_\perp=\vec p_{1\perp}+\vec p_{2\perp}$.  
{
Consequently, we can assume that the relative transverse momentum of the lepton pair $\vec P_\perp=(\vec p_{1\perp}-\vec p_{2\perp})/2 \sim p_{1\perp} \sim p_{2\perp}$ is much large than the total transverse momentum of lepton pair $\vec q_\perp=\vec p_{1\perp}+\vec p_{2\perp}$, and the lepton pair is almost back-to-back. It is widely referred to as the back-to-back correlation limit.}  Thus, we can use the relative transverse momentum of the lepton pair $P_\perp$ as the largest scale to factorize the process into two different parts, the hard part and the photon distribution part. The hard part corresponds to the hard interaction of lepton pair production, which depends on the hard momentum $ P_\perp$ and its azimuthal angle $\phi_P$. The photon distributions involve soft momentum scales, which are of the order of $q_\perp$ much less than $P_\perp$. It is widely referred to as the back-to-back correlation limit.
In this situation, one often needs to use the generalized equivalent photon approximation and the photon Wigner distribution to describe these quasi-real photons. The photon Wigner distribution, a quasi probability distribution function, includes the five-dimensional information of longitudinal momentum fraction, transverse momentum, and impart parameter of the photon, which is defined as~\cite{vonWeizsacker:1934nji,Williams:1934ad,Ji:1996ek, Belitsky:2000jk, Diehl:2001pm, Ji:2003ak, Belitsky:2003nz, Lorce:2011kd, Hatta:2016dxp, Klein:2020jom}
\begin{eqnarray}
\Gamma^{ij}(x,\vec  k_{T},\vec b_{\perp}) = \int \frac{d^2 \Delta_\perp}{(2\pi)^2}e^{i\Delta_\perp\cdot b_\perp}  \Gamma^{ij}(x,\vec  k_{T},\vec \Delta_\perp),
\end{eqnarray}
and $\Gamma^{ij}(x,\vec  k_{T},\vec \Delta_\perp)$ represents the so-called Generalized Transverse Momentum Dependent (GTMD) photon distributions in the momentum space. It reads 
\begin{eqnarray}
&&\Gamma^{ij}(x,\vec  k_{T}, \vec \Delta_\perp) = \int\frac{d\xi ^- d^2r_\perp}{(2\pi)^3}e^{ixP^+\xi^--ik_T\cdot r_\perp}\nonumber \\
&& \times \langle A,- \frac{\Delta_\perp}{2} | F^{+i}(0, \frac{r_\perp}{2}) F^{+j}(\xi^-, -\frac{r_\perp}{2})   |\frac{\Delta_\perp}{2},A  \rangle,
\end{eqnarray}
with the two dimensional polarization indices $i$ for the incoming photon in the scattering amplitude and the polarization indices $j$ for the photon in the conjugate amplitude. $k_{T}$ represents the transverse momentum of the photon, and $\Delta_\perp$ corresponds to the Fourier transform of $b_\perp$. Meanwhile, $x$ stands for the longitudinal momentum fractions of the photon per nucleon. Conventionally, the GTMD photon distributions can be parametrized into two parts:  the unpolarized photon distribution $xf_\gamma(x, \vec k_{T}, \vec \Delta_\perp)$ and the linearly-polarized photon distribution $xh_\gamma(x, \vec k_{T}, \vec \Delta_\perp)$. It is defined as
\begin{eqnarray}
\Gamma^{ij}(x,\vec  k_{T}, \vec \Delta_\perp) &=& \frac{\delta^{ij}}{2} xf_\gamma(x,  \vec k_{T}, \vec \Delta_\perp)\nonumber \\
&+& \left( \frac{k^i_+k^j_{-}}{\vec k_- \cdot \vec k_+}- \frac{\delta^{ij}}{2} \right)xh_\gamma(x, \vec k_{T}, \vec \Delta_\perp), 
\end{eqnarray}
with $k_{\pm}=k_T \pm \Delta_\perp/2$. In the correlation limit assumption, the dilepton production from the two-photon fusion process can be separated into two parts: the Wigner distribution and the hard scattering part. Thus, the factorization formula reads as follows
\begin{eqnarray}
&&\frac{d\sigma}{dy_1dy_2d^2q_\perp d^2P_\perp d^2b_\perp} =\int \Gamma_{ij}(x_{1},\vec  k_{1T},\vec b_{1\perp}) \otimes  \nonumber \\ 
 && \quad \quad \quad \Gamma_{kl}(x_{2},\vec k_{2T},\vec b_{2\perp}) \otimes \mathcal{H}^{ijkl}(\vec P_\perp). 
\label{eq:fac}
\end{eqnarray}
The two photon Wigner distributions, $\Gamma_{ij}$ and $\Gamma_{kl}$, contain the photon distribution information with the momentum scales significantly less than $ P_\perp$. The hard factor, $\mathcal{H}^{ijkl}$, represents the hard scale part and depends on the momentum imbalance of the lepton pair $ P_\perp$ and its angle $\phi_P$. Since the lepton mass $m_l$ is much smaller than $P_\perp$, we have neglected the lepton mass in the hard factor. The hard factor can be expressed as follows
\begin{eqnarray}
\mathcal{H}^{ijkl}(\vec P_\perp) & = & \frac{\alpha_{{\rm em}}^{2}}{\hat{s}^{2}} \left[O^{ijkl}(\vec P_{\perp})-4\Omega^{ijkl}(\vec P_{\perp}) \right]
,\label{eq:csm}
\end{eqnarray}
with the projection hard parts $O^{ijkl}$ and $\Omega^{ijkl}$ defined as
\begin{eqnarray}
O^{ijkl}(\vec P_{\perp})&=&2\left(\frac{\hat{u}}{\hat{t}}+\frac{\hat{t}}{\hat{u}}\right)\left(\delta^{ij}\delta^{kl}-\delta^{ik}\delta^{jl}+\delta^{il}\delta^{jk}\right),\\
\Omega^{ijkl}(\vec P_{\perp})&=&2\Pi^{ij}(\vec P_{\perp})\Pi^{kl}(\vec P_{\perp}) \nonumber \\
&&-\left(\delta^{il}\delta^{jk}+\delta^{ik}\delta^{jl}-\delta^{ij}\delta^{kl}\right),
\end{eqnarray}
where $\hat s$, $\hat u$ and $\hat t$ are the Mandelstam variables and $\Pi^{ij}(\vec P_{\perp})=2\frac{P_{\perp}^{i}P_{\perp}^{j}}{P_{\perp}^{2}}-\delta^{ij}$. 
The first term $O^{ijkl}$ is isotropic in terms of the angle $\phi_P$ as derived in Refs.~\cite{Vidovic:1992ik, Klein:2020jom}. {The second term, $\Omega^{ijkl}$, can generate azimuthal asymmetries between two linearly polarized photons, which is absent in the Ref~\cite{Klein:2020jom} due to the angular average of $\vec P_{\perp}$.} Given an arbitrary two-dimensional unit vector $\hat{n}$, it is interesting to note the following tensor contractions: $\hat{n}_i \hat{n}_j \Pi^{ij}  = \cos (2 \Delta \phi)$ and 
$\hat{n}_i \hat{n}_j \hat{n}_k \hat{n}_l \Omega^{ijkl}  = \cos (4\Delta \phi)$, 
where $\Delta \phi$ represents the azimuthal angle between $\vec{P}_\perp$ and $\hat{n}$. Integrating over the angle of impact parameter $\phi_b$, we find that the $\Omega^{ijkl}$ term projects the cross-section onto the $\cos(4\phi_P-4\phi_q)$ harmonics, reproducing the result derived in Ref.~\cite{Li:2019sin}. Similarly, after integrating over $q_\perp$, one obtains the $\cos(4\phi_P-4\phi_b)$ azimuthal asymmetry proposed in Ref.~\cite{Xiao:2020ddm} from the $\Omega^{ijkl}$ tensor. In general, various azimuthal angle asymmetries can arise from inner products of the three vectors $\vec{P}_\perp$, $\vec{q}_\perp$, and $\vec{b}_\perp$.

As discussed in Ref.~\cite{Klein:2020jom}, it is vital to demonstrate that this dilepton production cross-section always remains positive definite although the photon Wigner distribution may become negative. Let us rewrite the photon GTMD distribution as
\begin{equation}
\Gamma^{ij}(x_{1},\vec k_{1T},\vec \Delta_\perp) = \frac{Z^2  \alpha_{\rm em}}{\pi^2} k^i_{1\perp}{k^\prime}^j_{1\perp} \frac{F_A(k_{1})}{k_{1}^{2}} \frac{F_A(k_{1}^{\prime})}{{k^{\prime}}_{1}^{2}}, \,
\end{equation}
where $F_A(k_{1})$ is the elastic charge form factor, $Z$ is charge number and $m_p$ is the mass of proton with $k_{1}^2 = x_1 ^2m_p^2+k_{1\perp}^2$ and ${k^{\prime}}_{1}^{2} = x_1 ^2m_p^2+{k^{\prime}}_{1\perp}^{2}$. As detailed in the supplemental material, one finds
\begin{equation}
d\sigma \geq\left(\frac{Z^{2}}{\pi^{2}}\right)^{2}\frac{8\alpha_{{\rm em}}^{4}}{(2\pi)^{2}\hat{s}^{2}}(G^{12}+G^{21})(G^{12*}+G^{21*}),
\end{equation}
where $G^{ik}\equiv\int d^{2}\kaperp e^{i\kaperp\cdot b_\perp} F_A(k_1)F_A(k_2) |\kaperp||\kcperp|\hat{l}_{1}^{i}\hat{l}_{2}^{k}$. The $\hat{l}_{i}$ here refers to a unit vector, pointing the direction with the angle $\phi_{l_{i}}=\phi_{k_{i\perp}}-\phi_{P}$. Thus, it is clear that the cross-section is ensured to be positive definite.

\section{Phenomenology}
 Let us employ the unified factorization formalism to calculate various relevant observables, with a comprehensive phenomenological study available in the supplemental material.

First, the cross-section contains abundant angular correlation between the impact parameter and the transverse momentum of the final particles, and it can be decomposed as follows
\begin{eqnarray}
\sigma
&=& \sigma_0+ \sigma_{2qb} \cos (2\phi_q -2\phi_b)  + \sigma_{4qb} \cos (4\phi_q -4\phi_b)  \nonumber  \\
&& + \sigma_{4qP} \cos (4\phi_q -4\phi_P)     +\sigma_{4bP} \cos (4\phi_b -4\phi_P)   \nonumber  \\
&&
+ \sigma_{qbP}   \cos (2\phi_q +2\phi_b-4\phi_P)  +...,
\end{eqnarray}
where the coefficients $\sigma_{2qb}, \sigma_{4qb}, ...$ encode various information of the photon flux. As shown in Refs.~\cite{Li:2019yzy,Li:2019sin, Xiao:2020ddm, Klein:2020jom,Klein:2020fmr,Klusek-Gawenda:2020eja, Brandenburg:2020ozx, Shao:2022stc, Shao:2023zge}, the linearly polarized photons can generate a range of angular correlations either between $P_\perp$ and $q_\perp$ (including $\Delta \phi$ or $\alpha$ distributions at fixed $b_\perp$), or between $P_\perp$ and $b_\perp$. 

From the unified framework, one sees that there are a series of non-trivial angular correlations among $q_\perp$, $b_\perp$, and $P_\perp$.  {
We find that these anisotropies can be divided into two categories as shown in Tab. \ref{tab:aniso}. }

The first category, which does not depend on $P_\perp$, arises from the contraction between the $O^{ijkl}$ term of hard factor and the photon Wigner distributions. It leads to two sizable anisotropies, $2\langle \cos(2\phi_b-2\phi_q) \rangle$ and $2\langle \cos(4\phi_b-4\phi_q) \rangle$. {More precisely, when we analyze the cross-section in momentum space, these correlations can be generated by the phase factor  $e^{i \vec b_\perp \cdot \vec \Delta_\perp}$ and the linearly polarized photon distribution after contracting with the hard factor.}

The other category, which involves the $\cos(4\phi_P)$ correlation, is due to the contraction between the $\Omega^{ijkl}$ term and the photon Wigner distributions.  We can reproduce the two anisotropies, $2\langle\cos(4\phi_P-4\phi_q) \rangle$ and $2\langle\cos(4\phi_P-4\phi_b) \rangle$, as proposed earlier in Refs.~\cite{Li:2019yzy, Li:2019sin, Xiao:2020ddm}. Also, we find that there are additional novel angular correlations, such as $2\langle \cos (2\phi_q +2\phi_b-4\phi_P)  \rangle$, $2\langle \cos (6\phi_q -2\phi_b-4\phi_P)  \rangle$ and $2\langle \cos (6\phi_b -2\phi_q-4\phi_P)  \rangle$, etc.

\begin{table}[!ht]
\begin{tabular}{|c|l|}
\hline
category & anisotropies \\\hline
\multirow{2}{*}{1. $O\otimes xh_\gamma$} & $  \langle 2 \cos(2\phi_q-2\phi_b) \rangle$ \\
 & $ \langle 2 \cos(4\phi_q-4\phi_b) \rangle$   \\\hline
\multirow{7}{*}{2. $\Omega\otimes xh_\gamma$} & $ \langle 2 \cos(4\phi_P-4\phi_b) \rangle$\\
 &  $ \langle 2 \cos(4\phi_P-4\phi_q) \rangle$\\
 &  $ \langle 2 \cos(4\phi_P-2\phi_q-2\phi_b) \rangle$\\
 &  $ \langle 2 \cos(4\phi_P+2\phi_q-6\phi_b) \rangle$\\
 &  $ \langle 2 \cos(4\phi_P+2\phi_b-6\phi_q) \rangle$\\
 & $ \langle 2 \cos(4\phi_P+4\phi_q-8\phi_b) \rangle$\\
 & $ \langle 2 \cos(4\phi_P+4\phi_b-8\phi_q) \rangle$\\\hline
\end{tabular}
\caption{The table displays two categories of anisotropies: 1. The first two arise from the coupling of the \( O \)-term with the linearly polarized photon Wigner distribution. 2. Additional anisotropies result from the coupling of the \( \Omega \)-term with the same distribution.}
\label{tab:aniso}
\end{table}

\begin{figure*}[!ht]
\centering{
\includegraphics[width=0.8\textwidth]{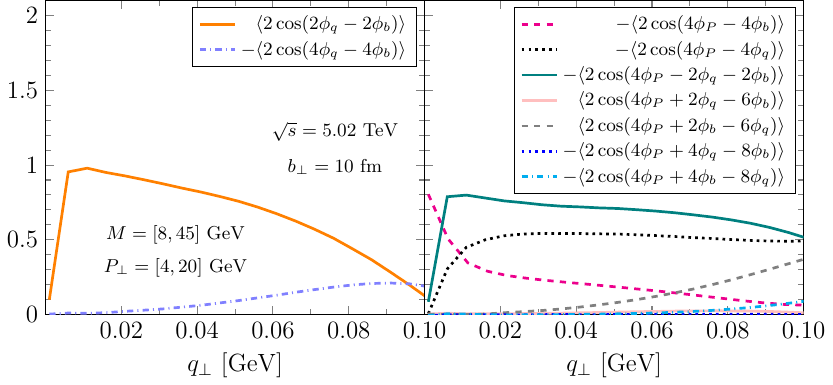}
}
\caption{
Lepton pair anisotropies as a function of $q_\perp$ with pair mass $M=8-45$ GeV, fixed $b_\perp=10$ fm and $P_\perp=4-20$ GeV at the LHC kinematics. The rapidities of the dilepton are integrated over the regions $[-1,1]$.
The left panel are the anisotropies from the first category $O\otimes xh_\gamma$, and the right panel are the anisotropies from the second category $\Omega\otimes xh_\gamma$.}
\label{fig:cos_LHC_q}
\end{figure*}

{
Fig. \ref{fig:cos_LHC_q} shows the anisotropies as a function of $q_\perp$ at the LHC energy. One may refer to Section 2.1 (Anisotropies) of the supplemental material for anisotropies as a function of $b_\perp$ and at RHIC energy, along with the detailed explanations. Since these various and sizeable anisotropies originate from the coupling of the hard factor with the linear-polarized photon, they effectively function as the a multi-dimensional imaging tool, helping us to study and detect the intrinsic properties of linear-polarized photons.}
It is interesting to note that the impact parameter vector $\vec{b}_\perp$ can be determined through the reaction plane of the hadronic events in peripheral collisions. Therefore, combined with the additional information provided by the momentum anisotropy of the produced hadrons in the same event, it is possible to measure all the aforementioned dilepton angular correlations among $q_\perp$, $P_\perp$, and $b_\perp$ in peripheral collisions at RHIC and the LHC. 

To provide a complete description of the most up-to-date data, we also need to include the incoherent contributions from the interior of the colliding nuclei. When the transverse momentum of incoming photons is less than $1/R_A$ with $R_A$ the nuclear radius, one can view the nucleus as a coherent charge source of photons. While incoming photons emerge incoherently from the interior of the heavy nuclei, they often carry much larger transverse momentum as they resolve the nucleus' internal sub-structure. This scenario is often accompanied by the dissociation of these nuclei. The STAR, CMS and ATLAS Collaborations have recently published a substantial amount of data related to neutron emissions and nuclear break-ups in UPCs~\cite{STAR:2019wlg, ATLAS:2020epq, CMS:2020skx, ATLAS:2022srr}. In the following, we examine the role of incoherent contributions in events involving neutron emissions. 

In the $0n0n$ scenario in which no neutron is detected, the heavy nucleus remains intact. The primary contribution comes from the coherent photons. As substantiated in previous studies~\cite{Klein:2018fmp, Klein:2020jom, Shao:2023zge}, the large transverse momentum imbalance of the lepton pair induced by the final-state soft multi-photon radiation can be effectively addressed through the Sudakov resummation. The detailed implementation is provided in the supplemental material. On the other hand, for the $XnXn$ cases where neutron emissions are observed, the heavy nucleus is considered disintegrated, indicating the presence of not only the final-state soft multi-photon radiation but also incoherent interactions in these events. 
Therefore, taking into account photons come from different sources, we modified the GTMD photon distribution $\Gamma^{ij}(x,k_{T},\Delta)$ as follows
\begin{eqnarray}
\Gamma^{ij}(x, \vec k_{T},\vec \Delta_\perp)
&=& \Gamma_A^{ij}(x, \vec k_{T}, \vec  \Delta_\perp) +\Gamma_p^{ij}(x, \vec k_{T},\vec \Delta_\perp) \nonumber \\
&&+
\Gamma_q^{ij}(x, \vec k_{T},\vec  \Delta_\perp).
\end{eqnarray}
The first term $ \Gamma_A^{ij}(x, \vec k_{T}, \vec \Delta_\perp) $ represents the coherent contribution from the heavy nucleus, acting as a whole, as the photon source. The second term $ \Gamma_p^{ij}(x, \vec k_{T}, \vec \Delta_\perp) $ signifies that the proton inside the nucleus plays the role of the individual photon source, which is given by 
\begin{equation}
\Gamma_p^{ij}(x, \vec k_{T}, \vec \Delta_\perp)= Z \frac{\alpha_{\rm em}}{\pi^2} k^i_{\perp}{k^\prime}^j_{\perp}  \frac{F_p(k)}{k^{2}} \frac{F_p(k^{\prime})}{{k^{\prime}}^{2}},
\label{eq:xfp2}
\end{equation}
where the $Z$ factor arises from sum over all protons in the nucleus and $F_p(k)$ stands for the proton charge form factor. The last term corresponds to the incoherent contribution originated from quarks inside nucleons. By treating quarks as point-like particles, then the corresponding GTMD photon distribution $\Gamma_q^{ij} (x_, \vec k_{T}, \vec \Delta_\perp) $ reads {
\begin{equation}
\Gamma_q^{ij} (x, \vec k_{T}, \vec \Delta_\perp) = \left( \sum_{q/p} e_q^2+\sum_{q/n} e_q^2 \right) \frac{ \alpha_{\rm em}}{\pi^2} k^i_{\perp}{k^\prime}^j_{\perp}
 \frac{1}{k^{2}} \frac{1}{{k^{\prime}}^{2}},
\end{equation}
where $e_q$ is the quark charge number. The summed factor $\sum_{q/p} e_q^2+\sum_{q/n} e_q^2$ arise from sum over all quarks from the protons and neutrons inside nucleus.} In the experimental measurements, the maximum value of transverse momentum imbalance is approximately $1.2$ GeV, which is already much larger than the typical confinement scale. Thus, at a scale around $1$ GeV, dilepton production is capable of resolving quarks within the proton, although fluctuations inside the nucleon remain relatively insignificant. Consequently, for the sake of simplifying the calculation, we utilize the valence quark model to outline the quark distribution within both the proton and the neutron. We assume that these three quarks, two up quarks and one down quark, share the same momentum within the proton. Similarly, two down quarks and one up quark share the same momentum within the neutron. The above approximation can be improved by following a recent study of the photon parton distribution in the proton~\cite{Manohar:2016nzj,Manohar:2017eqh}, where a rigorous connection to the proton structure functions played an important role. We expect a similar treatment can be made for the above incoherent contribution to the photon Wigner distribution. We will leave this issue for a later publication.

\begin{figure}[!ht]
\includegraphics[width=0.99\linewidth]{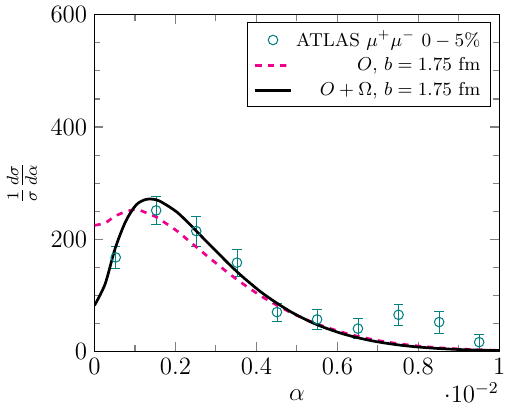}
\includegraphics[width=0.99\linewidth]{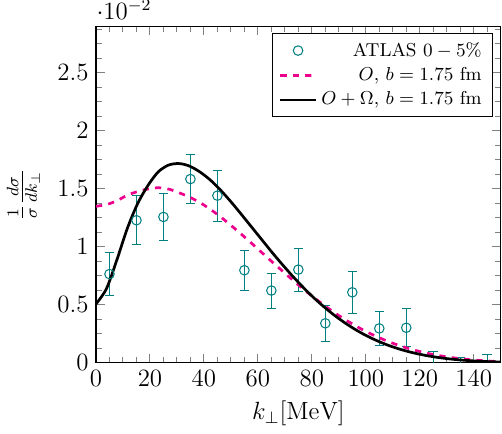}
\caption{Comparison the numerical results with the ATLAS measurement of the dimuon pair acoplanarity $\alpha$ (upper) distribution and $k_\perp$ (lower) distribution in central collisions~\cite{ATLAS:2022yad}.}
\label{fig:atlas}
\end{figure} 

\section{Comparing with data}
\label{sec:comp}
In our numerical calculation, we follow the STARlight~\cite{Klein:2016yzr} and use the nucleus form factor\cite{Davies:1976zzb}
\begin{equation}
F_A(k)=\frac{3}{(kR_A)^3}\frac{\left[\sin(kR_A)-kR_A\cos(kR_A)\right]}{1+a^2k^2},
\end{equation}
with $a=0.71$ fm, $R_A$ nucleus radius and $A$ the number of nucleons in the nucleus. We set $R_A=6.6$ fm for the lead nucleus. For the proton, we employ the dipole form factor~\cite{Andivahis:1994rq,Klein:2003vd}
\begin{equation}
F_p(k) =\frac{1}{[(k^2/0.71 \text{ GeV}^2)+1]^2}.
\end{equation}
All of our numerical results and detailed comparisons with RHIC and LHC data can be found in the supplemental material. Here, we present a few selected results for illustration.

As depicted in Fig.~\ref{fig:atlas}, we compare our results with the $\alpha$ and $k_\perp$ distribution data from the ATLAS Collaboration~\cite{ATLAS:2022yad} in the $0-5\%$ centrality interval. The acoplanarity $\alpha$ is defined by $\alpha =1- \Delta \phi/\pi$, where $\Delta \phi$ denotes the angular difference between two leptons, and $k_\perp$ is defined as $k_\perp=\pi\alpha P_\perp$. The magenta dashed line represents the computation derived from the first term $O^{ijkl}$ of the hard factor, whereas the black solid line indicates the calculation resulting from both the first term $O^{ijkl}$ and the second term $\Omega^{ijkl}(\vec P_\perp)$.
In the small $\alpha$ and small $k_\perp$ region, there is a notable gap between the result computed from the first term $O^{ijkl}$ of the hard factor and the experimental data. However, after including the second term $\Omega^{ijkl}(\vec P_\perp)$, our calculation can reproduce the dip shape observed in the experiment and perfectly aligns with the ATLAS experimental data, thereby highlighting the importance of the $\Omega^{ijkl}(\vec P_\perp)$ term in describing the dip in the extremely small $\alpha$ and $k_\perp$ region. Similar numerical results can also be found in the calculations in Refs.~\cite{Zha:2018tlq, Li:2019yzy, Li:2019sin, Wang:2021kxm, Wang:2022gkd, Shao:2023zge}. As $\alpha$ increases, we find that the contribution of the $\Omega^{ijkl}(\vec P_\perp)$ term tends to diminish, and the $O^{ijkl}$ term takes the dominant contribution. The $k_\perp$ distribution also shows the same trend.

\begin{figure}[!ht]
\includegraphics[width=.99\linewidth]{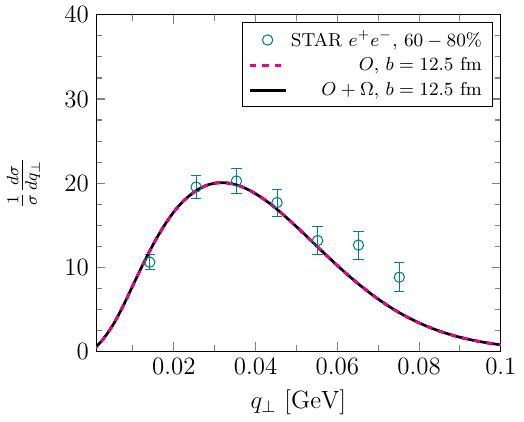}
\caption{Comparisons of our results with the STAR experimental data~\cite{STAR:2019wlg} in the dilepton pair transverse momentum imbalance  $q_\perp$ distribution at $60-80\%$ central collisions. } 
\label{fig:star_qt}
\end{figure}

{
Additionally, as shown in Fig.~\ref{fig:star_qt}, our results can provide a good description of the dilepton $q_\perp$ distribution at 60-80\% centrality compared with the STAR experimental data~\cite{STAR:2019wlg}.
One can find that the second term $\Omega^{ijkl}(\Vec{P}_\perp)$ of the hard factor does not have a contribution to the $q_\perp$ distribution. 
Since the $q_\perp$ distribution is angular average observable, and the azimuthal angle of the $\phi_P$ is integrated out, the term $\Omega^{ijkl}(\Vec{P}_\perp)$ vanishes, and only the average term $O^{ijkl}$ contributes.}

\begin{figure*}[!ht]
\includegraphics[width=.45\linewidth]{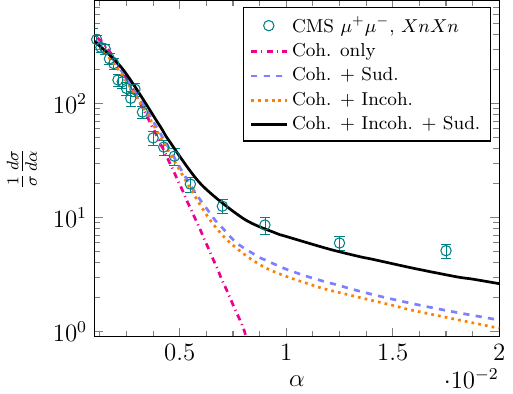}
\includegraphics[width=.45\linewidth]{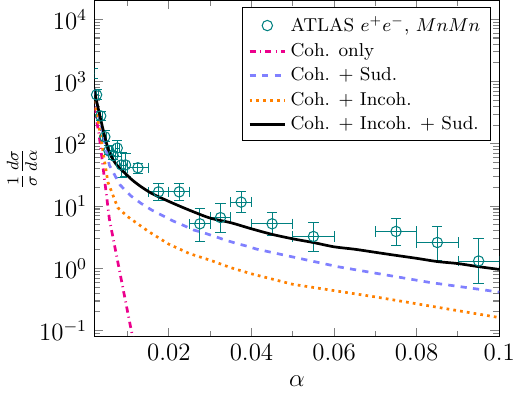}
\caption{Comparisons of the numerical results with the CMS $XnXn$ data~\cite{CMS:2020skx} and the ATLAS $MnMn$ data~\cite{ATLAS:2022srr}.} 
\label{fig:ne_em}
\end{figure*} 

We compare our numerical calculations for the $\alpha$ distribution involving neutron emissions with the CMS $XnXn$ and ATLAS $MnMn$ data~\cite{CMS:2020skx, ATLAS:2022srr} in Fig.~\ref{fig:ne_em}. In the CMS experiment, $Xn$ represents the emission of two or more neutrons, whereas $Mn$ denotes the emission of one or more neutrons in the ATLAS experiment. The magenta dash-dotted line stands for the coherent calculation without the Sudakov effect, and the blue dashed line corresponds to the coherent calculation with the Sudakov effect. In addition, the orange dotted line represents the sum of the coherent and incoherent contribution without the Sudakov effect. At last, the solid line includes both incoherent contribution and the Sudakov effect.   

All four calculations agree in the small-$\alpha$ region, meaning that the coherent term originating from the nucleus acting as the photon source is the dominant contribution, and the incoherent contribution and Sudakov effect are negligible in this region. In the large-$\alpha$ region, both the Sudakov effect and the incoherent contribution are important mechanisms leading to the final $q_\perp$ broadening. However, neither of these mechanisms alone can accurately describe the data with neutron emissions. When both mechanisms are considered, we can obtain the enhanced broadening effect and find good agreements with the CMS and ATLAS data. This implies that the final $q_\perp$ broadening in the large-$\alpha$ region receive contributions from both the Sudakov effect and the incoherent contribution. In the small-$\alpha$ region, the corresponding transverse momentum imbalance, $q_\perp \sim \pi \alpha P_{\perp}$, is less than $30$ MeV. As a result, dileptons cannot penetrate or ``see" the inner structure of heavy nuclei, allowing the heavy nucleus to be perceived as a whole entity radiating photons. However, as $\alpha$ increases, $q_\perp$ can rise up to $1\sim 3$ GeV. This enables the dilepton to resolve the inner structure of heavy nuclei, such as protons and quarks. Consequently, quarks and protons begin to act as incoherent photon sources in the large $\alpha$ region.

\section{Conclusion}
\label{sec:conc}

For lepton pair production via the photon fusion process in heavy-ion collisions, we introduce a novel unified factorization framework that decomposes the cross-section into the soft photon Wigner distribution and the photon fusion hard factor—including the angular average of the $\vec{P}_\perp$ term and the anisotropy term—allowing for an independent analysis of each component.

Anisotropies serve as the `golden' observables for studying the properties of linearly polarized photons in both experimental measurements and theoretical research. Previous studies have investigated two sizeable anisotropies, $\cos(4P_\perp - 4q_\perp)$ and $\cos(4P_\perp - 4b_\perp)$, which arise from the coupling between linearly polarized photons and the $\Omega$ term of the hard factor. Our unified framework enables us to extract various correlations, and for the first time, we reveal that sizeable anisotropies originate not only from the coupling of linearly polarized photons with the $\Omega$ term but also from their coupling with the $O$ term. Future experimental studies at RHIC and the LHC can explore these anisotropies to further probe the intrinsic properties of linearly polarized photons in heavy-ion collisions.  

Additionally, our study shows that the photon source is not solely from the heavy nucleus as a whole but also from the protons inside the nucleus and the substructure of the nucleons (the quarks), which have not been previously considered in other studies. Considering events accompanied by neutron emissions measured at the LHC, we show that the incoherent contributions are significant in the large-$\alpha$ region.

In comparison with previous studies, we summarize the key new developments in our work:
\begin{itemize}  
\item Building upon the previous GTMD framework, which only included the angular average of the $P_\perp$ contribution, we demonstrate that the anisotropic contributions from the hard part can also be factorized, leading to a fully unified factorization framework.  
\item In addition to the two dilepton anisotropies previously studied, we identify an entirely new class of anisotropies that can be used to probe the intrinsic structure of the proton.  
\item While previous studies considered only coherent contributions, we incorporate incoherent contributions naturally into the photon Wigner distribution for the first time, which provides a potential explanation for the acoplanarity distribution with neutron emission in the large-$\alpha$ region.  
\item We perform a comprehensive analysis and successfully describe a wide range of relevant data from the STAR, CMS, and ATLAS Collaborations, covering both low and high $q_\perp$ regions.  
\end{itemize}

Finally, this unified factorization formula, developed in this study, can offer novel perspectives for exploring the factorization and production mechanisms of other processes in UPCs.

\section*{Acknowledgements}
We thank Feng Yuan, Jian Zhou, Cheng Zhang, Ya-Jin Zhou and Shi Pu for useful inputs and discussions. This work is partly supported by the CUHK-Shenzhen under grant No. UDF01001859. Y. Shi is supported by the China Postdoctoral Science Foundation under Grant No. 2022M720082. S.Y. Wei is supported by the Shandong Province Natural Science Foundation under grant No.~2023HWYQ-011 and the Taishan fellowship of Shandong Province for junior scientists.

\begin{widetext}

\section*{Supplemental material}
\resumetoc

\begin{center}
Yu Shi, Lin Chen, Shu-Yi Wei, Bo-Wen Xiao
\end{center}

\tableofcontents

\vspace{1.cm}

We provide additional details in the supplemental material. In Sec.~\ref{sec::fac}, the generalized factorization framework for the dilepton production via the photon-photon fusion process in ultra-peripheral heavy-ion collisions is presented. Sec.~\ref{sec::phe} discusses several topics concerning the phenomenological studies. In Sec.~\ref{sec:com}, we present additional numerical results compared with the experimental data. 

\subsection{The Generalized Factorization formalism}
\label{sec::fac}

\begin{figure}[!h]
\center{
\includegraphics[height=6.5cm]{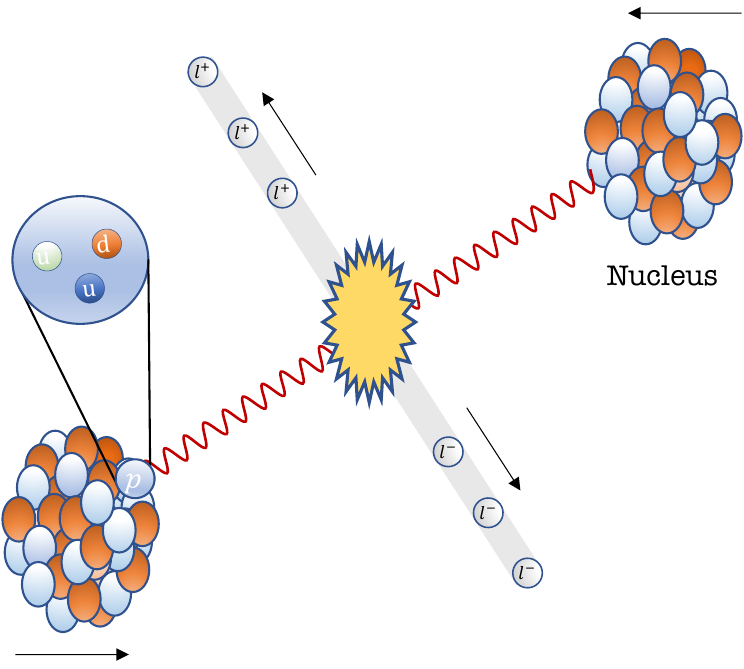}
\qquad
\qquad
\includegraphics[height=6.5cm]{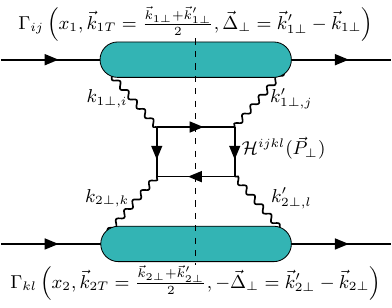}
}
\caption{Left: The cartoon depiction of the lepton pair production from the photon-photon fusion process in heavy-ion collisions. Right: The Feynman diagram illustrating the generalized TMD factorization for the $\gamma(k_1)+\gamma(k_2) \rightarrow l^+(p_{1}) + l^-(p_{2})$ process in heavy-ion collisions.}  
\label{fig:lep}
\end{figure}

The left figure in Fig.~\ref{fig:lep} shows the cartoon of the dilepton production from electromagnetic interactions in heavy-ion collisions: two quasi-real photons, each radiated from one of the two fast-moving nuclei at a relative distance $b_\perp$, undergo fusion to generate a pair of leptons. These leptons are almost back-to-back, each with transverse momentum $p_{1\perp}$ and $p_{2\perp}$, and are produced at the rapidity $y_1$ and $y_2$, respectively. The collision point, where these leptons are produced, is situated at distances $b_{1\perp}$ and $b_{2\perp}$ from the two heavy nuclei.

In the following calculations, we assume that the relative transverse momentum of the lepton pair $\vec P_\perp=(\vec p_{1\perp}-\vec p_{2\perp})/2 \sim p_{1\perp} \sim p_{2\perp}$ is much large than the total transverse momentum of lepton pair $\vec q_\perp=\vec p_{1\perp}+\vec p_{2\perp}$. Thus, we can use the relative transverse momentum of the lepton pair $P_\perp$ as the largest scale to factorize the process into two different parts, the hard part and the photon distribution part. The hard part corresponds to the hard interaction of lepton pair production, which depends on the hard momentum $ P_\perp$ and its azimuthal angle $\phi_P$. The photon distributions involve soft momentum scales, which are of the order of $q_\perp$ much less than $P_\perp$. 

The study of photon distributions within the nucleus has a long history. Initially, the equivalent photon approximation (EPA), also known as the Weizs\"acker-Williams (WW) approximation method~\cite{vonWeizsacker:1934nji,Williams:1934ad}, is the most commonly used method for addressing photon distribution. In this method, photons are treated as quasi-real particles, and by calculating the lowest-order contributions in QED, one can obtain a photon distribution with low virtuality and transverse momentum. However, the original approximation does not account for both the transverse momentum and the impact parameter of the photon. Recently, inspired by the implementation of the Wigner distribution in QCD studies~\cite{Ji:2003ak, Belitsky:2003nz, Lorce:2011kd, Hatta:2016dxp}, the generalized equivalent photon approximation including the transverse momentum and the impact parameter dependence has been proposed to parameterize the photon distribution from high energy nuclei, which is also known as the photon Wigner distribution~\cite{Li:2019sin, Klein:2020jom}. Also, one can define the Generalized Transverse Momentum Dependent (GTMD) distribution by Fourier transforming the impact parameter dependence into the momentum space.

As illustrated in the right figure of Fig.~\ref{fig:lep}, the generalized factorization formula for the process  following the GTMD in the correlation limit can be cast into
\begin{eqnarray}
\frac{d\sigma}{dy_{1}dy_{2}d^{2}P_{\perp}d^{2}q_{\perp}d^{2}b_{\perp}} & = & \frac{1}{(2\pi)^4} \int d^{2}b_{1\perp}d^{2}b_{2\perp}\delta^{2}(\vec b_{\perp}-\vec b_{1\perp}+\vec b_{2\perp})\int d^{2}k_{1T}d^{2}k_{2T} \int d^{2}\Delta_{1\perp}d^{2}\Delta_{2\perp} e^{-ib_{1\perp}\cdot\Delta_{1\perp}-ib_{2\perp}\cdot\Delta_{2\perp}} \nonumber \\
&& \times \Gamma_{ij}(x_{1},\vec k_{1T},\vec  \Delta_{1\perp})\Gamma_{kl}(x_{2},\vec k_{2T}, \vec \Delta_{2\perp})\mathcal{H}^{ijkl}(\vec P_\perp) \delta^{(2)}(\vec q_{\perp}-\vec k_{1T}-\vec k_{2T}),
\label{eq:1}
\end{eqnarray}
where $\mathcal{H}^{ijkl}$ stands for the hard factor, $\Gamma_{ij}$ and $\Gamma_{kl}$ represent the 5-dimensional GTMD distribution with the polarization indices $i$ and $k$ of incoming two photons in the scattering amplitude and the polarization indices $j$ and $l$ of photons in conjugate amplitude. These two-dimensional indices run from $1$ to $2$. To reconstruct the kinematics of the incoming photons, we write the longitudinal momentum fraction of photon per nucleon as $x_1=\text{max}[p_{1\perp},p_{2\perp}](e^{y_1}+e^{y_2})/\sqrt{s} $ and $x_2=\text{max}[p_{1\perp},p_{2\perp}] (e^{-y_1}+e^{-y_2})/\sqrt{s} $. $b_{1\perp}$ and $b_{2\perp}$ stand for the distance of the photon and its corresponding nucleus. $\vec k_{1T}=( \vec k_{1\perp}+\vec k_{1\perp}^{\prime}  )/2$ and $\vec k_{2T}=( \vec k_{2\perp}+\vec k_{2\perp}^{\prime}  )/2$ are transverse momenta of two photons, where $ k_{i\perp}$ and $ k_{i\perp}^{\prime}$ are the transverse
momenta of incoming photons in the amplitude and conjugate amplitude, respectively. According the momentum conservation, these momenta satisfy $\vec{q}_\perp= \vec k_{1\perp}+ \vec k_{2\perp}=\vec k_{1\perp}^{\prime}+\vec k_{2\perp}^{\prime}$.
$\vec \Delta_{1\perp}=\vec k_{1\perp}^{\prime}-\vec k_{1\perp}$ and $ \vec \Delta_{2\perp}=\vec k^{\prime}_{2\perp}-\vec k_{2\perp}$ are defined as the momentum difference for the nucleus state in the non-forward scatterings and the momentum conservation implies $\vec \Delta_{1\perp}+\vec \Delta_{2\perp}=0$. Thus we define $\vec\Delta_\perp = \vec\Delta_{1\perp} =-\vec \Delta_{2\perp}$ for convenience. It is important to note that the momenta from the scattering amplitude and conjugate amplitude only become identical in forward scattering with $\Delta_{1\perp} =\Delta_{2\perp} =0$. By integrating over the two delta functions, the cross-section can be rewritten as
\begin{eqnarray}
\frac{d\sigma}{dy_{1}dy_{2}d^{2}P_{\perp}d^{2}q_{\perp}d^{2}b_{\perp}} & = & \frac{1}{(2\pi)^{2}}\int d^{2}k_{1\perp}d^{2}k_{1\perp}^{\prime}e^{ib_{\perp}\cdot\Delta_\perp}\Gamma_{ij}(x_{1},\vec k_{1T},\vec  \Delta_\perp)\Gamma_{kl}(x_{2},\vec k_{2T},-\vec \Delta_\perp)\mathcal{H}^{ijkl}(\vec P_\perp),
\label{eq:11}
\end{eqnarray}
with $\vec \Delta_\perp=\vec k_{1\perp}^{\prime}-\vec k_{1\perp}=\vec k_{2\perp}-\vec k_{2\perp}^{\prime}$. By using the definition
\begin{equation}
\Gamma_{ij}(x,\vec k_T,\vec \Delta_\perp)=\int d^2b_\perp e^{-i\vec b_\perp\cdot\vec \Delta_\perp} \Gamma_{ij}(x, \vec k_T ,\vec b_\perp),
\end{equation}   
where $b_{\perp}$ is the distance of the photon and center of the nucleus, one can convert the GTMD photon distribution into the photon Wigner distribution via the Fourier transform. Then, the generalized factorization formula involving the Winger photon distribution becomes
\begin{eqnarray}
\frac{d\sigma}{dy_{1}dy_{2}d^{2}p_{1\perp}d^{2}p_{2\perp}d^{2}b_{\perp}} & = & \int d^{2}b_{1\perp}d^{2}b_{2\perp}\delta^{2}(\vec b_{\perp}-\vec b_{1\perp}+\vec b_{2\perp})\int d^{2}k_{1T}d^{2}k_{2T}  \nonumber \\
 &  & \times\Gamma_{ij}(x_{1},\vec  k_{1T},\vec b_{1\perp})\Gamma_{kl}(x_{2},\vec k_{2T},\vec b_{2\perp})\mathcal{H}^{ijkl}(\vec P_\perp)\delta^{(2)}(\vec p_{1\perp}+\vec p_{2\perp}-\vec k_{1T}-\vec k_{2T}).
\end{eqnarray}
In the meantime, with the definition 
\begin{equation}
\Gamma_{ij}(x,\vec k_{T},\vec b_{\perp}) =  \int d^{2}r_{\perp}e^{-i\vec k_{T}\cdot \vec r_{\perp}}\Gamma_{ij}(x,\vec r_{\perp},\vec b_{\perp}),
\end{equation}
where $r_\perp$ represents photon position difference between the amplitude and conjugate amplitude.
The cross-section in the full coordinate space can be written as 
\begin{eqnarray}
\frac{d\sigma}{dy_{1}dy_{2}d^{2}P_{\perp}d^{2}q_{\perp}d^{2}b_{\perp}}=\int d^{2}b_{1\perp}d^{2}r_{\perp}e^{-iq_{\perp}\cdot r_{\perp}}\Gamma_{ij}(x_{1},\vec  r_{\perp},\vec  b_{1\perp})\Gamma_{kl}(x_{2},\vec r_{\perp},-\vec b_{\perp}+\vec b_{1\perp})\mathcal{H}^{ijkl}(\vec P_\perp).
\end{eqnarray}

In order to derive the hard factor $\mathcal{H}^{ijkl}$, one needs to carry out the Feynman graph calculation for $\gamma \gamma \to l^+l^-$ with open indices for all four photon polarizations. After some tedious tensor manipulations, one can eventually find
\begin{eqnarray}
\mathcal{H}^{ijkl}(\vec P_\perp)  & = & \frac{\alpha_{\rm em}^2}{\hat s ^2} \Bigg\{   2\frac{(m_{l}^{4}-\hat{t}\hat{u})(6m_{l}^{4}-4m_l^{2}\hat{t}-4m_{l}^{2}\hat{u}+\hat{t}^{2}+\hat{u}^{2})}{(m_{l}^{2}-\hat{t})^{2}(m_{l}^{2}-\hat{u})^{2}}\left(\delta^{ij}\delta^{kl}-\delta^{ik}\delta^{jl}+\delta^{il}\delta^{jk}\right)\nonumber \\
 &  & \qquad \quad +\frac{4m_{l}^{4}(2m_{l}^{2}-\hat{t}-\hat{u})^{2}}{(m_{l}^{2}-\hat{t})^{2}(m_{l}^{2}-\hat{u})^{2}}\left(-\delta^{ij}\delta^{kl}+\delta^{ik}\delta^{jl}+\delta^{il}\delta^{jk}\right)\nonumber \\
 &  &   \qquad \quad  -\frac{2m_{l}^{2}(\hat{t}+\hat{u})(2m_{l}^{2}-\hat{t}-\hat{u})^{2}}{(m_{l}^{2}-\hat{t})^{2}(m_{l}^{2}-\hat{u})^{2}}\left(\delta^{ij}\delta^{kl}+\delta^{ik}\delta^{jl}-\delta^{il}\delta^{jk}\right)\nonumber \\
 &  &   \qquad \quad  -8m_{l}^{2}\frac{(2m_{l}^{2}-\hat{t}-\hat{u})(m_{l}^{4}-\hat{t}\hat{u})}{(m_{l}^{2}-\hat{t})^{2}(m_{l}^{2}-\hat{u})^{2}}\left[\delta^{ij}\Pi^{kl}(\vec P_{\perp})+\Pi^{ij}(\vec P_{\perp})\delta^{kl}\right]\nonumber \\
 &  &   \qquad \quad   -4\frac{(m_{l}^{4}-\hat{t}\hat{u})^{2}}{(m_{l}^{2}-\hat{t})^{2}(m_{l}^{2}-\hat{u})^{2}}\Omega^{ijkl}(\vec P_{\perp})  \Bigg\},
\end{eqnarray}
where $\hat s$, $\hat u$ and $\hat t$ are the Mandelstam variables. Inside the curly brackets, the first three terms are the isotropic contributions which are independent of the orientation of $\vec P_\perp$, the fourth term gives rise to the $\cos(2\phi_{P})$ angular correlation, and the last term can generate the $\cos(4\phi_{P})$ angular correlation. The tensors $\Pi^{ij}(\vec P_{\perp})$ and $\Omega^{ijkl}(\vec P_\perp)$ are defined as $\Pi^{ij}(\vec P_{\perp})=2\frac{P_{\perp}^{i}P_{\perp}^{j}}{P_{\perp}^{2}}-\delta^{ij}$ and $\Omega^{ijkl}(\vec P_{\perp})=2\Pi^{ij}(\vec P_{\perp})\Pi^{kl}(\vec P_{\perp})-\left(\delta^{il}\delta^{jk}+\delta^{ik}\delta^{jl}-\delta^{ij}\delta^{kl}\right)$. They can project the cross-sections onto $\cos(2\phi_{P})$ and $\cos(4\phi_{P})$ azimuthal angular correlations, respectively. For example, with four arbitrary two-dimensional unit vector $\hat{n}_{1,2,3,4}$, one can obtain
\begin{eqnarray}
&& \Pi^{ij} \hat{n}_{1i} \hat{n}_{2j} = \cos ( \phi_1 +\phi_2- 2 \phi_P) \, , \\
&& \Omega^{ijkl} \hat{n}_{1i} \hat{n}_{2j} \hat{n}_{3k} \hat{n}_{4l} = \cos (\phi_1+ \phi_2 +\phi_3+\phi_4-4\phi_P),
\end{eqnarray}
where $\phi_1$ represents the azimuthal angle orientation of $\hat{n}_1$, and similarly for other azimuthal angles.

If we average over the angle of impact parameter $b_\perp$ and project $\cos(2\phi_{P})$ terms into $\cos(2\phi_P-2\phi_q)$ and $\cos(4\phi_{P})$ term into $\cos(4\phi_P-4\phi_q)$, we can reproduce the result derived in Refs.~\cite{Shao:2022stc}. Since the contribution of lepton mass is power suppressed by factors of $m_l^2/P^2_\perp$, we neglect the mass of lepton effect in the hard factor. Then, the hard factor can be significantly simplified and it reads
\begin{eqnarray}
\mathcal{H}^{ijkl}(\vec P_\perp) & = & \frac{\alpha_{{\rm em}}^{2}}{\hat{s}^{2}} \left[O^{ijkl}-4\Omega^{ijkl}(\vec P_{\perp}) \right], \label{hard2}
\end{eqnarray}
with
\begin{eqnarray}
O^{ijkl}&=&2\left(\frac{\hat{u}}{\hat{t}}+\frac{\hat{t}}{\hat{u}}\right)\left(\delta^{ij}\delta^{kl}-\delta^{ik}\delta^{jl}+\delta^{il}\delta^{jk}\right),\\
\Omega^{ijkl}(\vec P_{\perp})&=&2\Pi^{ij}(\vec P_{\perp})\Pi^{kl}(\vec P_{\perp})-\left(\delta^{il}\delta^{jk}+\delta^{ik}\delta^{jl}-\delta^{ij}\delta^{kl}\right).
\end{eqnarray}
The terms related to $\cos(2\phi_{P})$ are proportional to the lepton's mass, thus they vanish in the massless limit. In the remaining terms, the $O^{ijkl}$ term represents the isotropic component with respect to $P_\perp$ and coincides with the angle-averaged $\phi_P$ term as discussed in Ref.~\cite{Klein:2020jom}. This term emerges from the contributions of both the unpolarized and the linearly-polarized photon distributions. The $\Omega^{ijkl}(\vec P_{\perp})$ term, which originates from the interaction between two linearly-polarized photons, will generate the $\cos(4\phi_{P})$ angular correlations in the following calculations.

In general, the GTMD photon distribution can be parametrized as follows
\begin{equation}
\Gamma^{ij}(x,\vec k_{T}, \vec \Delta_\perp)= \frac{\delta^{ij}}{2}xf_\gamma(x,\vec k_T, \vec \Delta_\perp) +\left( \frac{k^i_+k^j_{-}}{\vec k_- \cdot \vec k_+}- \frac{\delta^{ij}}{2} \right)xh_\gamma(x,\vec k_T, \vec \Delta_\perp), 
\end{equation}
with $\vec k_{\pm}=\vec k_\perp \pm \vec \Delta_\perp/2$. The first term represents the unpolarized photon distribution and the second term stands for linearly-polarized photon distribution. We can assume that the  unpolarized photon distribution is the same as the linearly-polarized photon distribution, and the GTMD photon distribution can be parametrized as
\begin{equation}
\Gamma^{ij}(x_{1},\vec k_{1T},\vec \Delta_\perp) = \frac{Z^2 \alpha_{\rm em}}{\pi^2} k^i_{1\perp}{k^\prime}^j_{1\perp} \frac{F_A(k_{1})}{k_{1}^{2}} \frac{F_A(k_{1}^{\prime})}{{k^{\prime}}_{1}^{2}},
\end{equation}
where $F_A(k_{1})$ is the normalized elastic charge form factor for the nucleus, $Z$ is charge number of nucleus and $m_p$ is the mass of proton with $k_{1}^2 = x_1 ^2m_p^2+k_{1\perp}^2$ and ${k^{\prime}}_{1}^{2} = x_1 ^2m_p^2+{k^{\prime}}_{1\perp}^{2}$. Now, the cross-section also can be written as 
\begin{eqnarray}
\frac{d\sigma}{dy_{1}dy_{2}d^{2}P_{\perp}d^{2}q_{\perp}d^{2}b_{\perp}} 
 & = & \frac{1}{(2\pi)^{2}} \frac{2\alpha_{{\rm e}}^{4}}{\hat{s}^{2}}  \left(\frac{Z^{2}}{\pi^{2}}\right)^{2}\int d^{2}k_{1\perp}d^{2}k_{1\perp}^{\prime}e^{ib_{\perp}\cdot\Delta_\perp}\frac{F_A(k_{1})}{k_{1}^{2}}\frac{F(k_{2})}{k_{2}^{2}}\frac{F_A(k_{1}^{\prime})}{{k^{\prime}}_{1}^{2}}\frac{F_A(k_{2}^{\prime})}{{k^{\prime}}_{2}^{2}}\kaperp\kbperp\kcperp\kdperp\nonumber \\
 &  & \times\left [ \left( \frac{\hat u}{\hat t}+\frac{\hat t}{\hat u}\right) \cos(\phi_{\kaperp}-\phi_{\kbperp}+\phi_{\kcperp}-\phi_{\kdperp})-  2\cos(\phi_{\kaperp}+\phi_{\kbperp}+\phi_{\kcperp}+\phi_{\kdperp}-4\phi_{P}) \right].\nonumber \\
 \label{eq:cos}
\end{eqnarray}
From the above equation, by averaging over the impact parameter angle $b_\perp$ and projecting the $\cos(4\phi_{P})$ term into $\cos(4\phi_P-4\phi_q)$ direction, we can reproduce the results derived in Refs.~\cite{Li:2019sin}. In the coordinate space, if we integrate over $q_{\perp}$ and $\phi_q$ and implement the delta function track in the anisotropy term, it is straightforward to find that the above equation is equal to the results given by Eq.~(5) in Ref.~\cite{Xiao:2020ddm}. One can find that all anisotropic terms with $\phi_q$ disappear, and only $ \cos (4\phi_b -4\phi_P) $ remains.

The cross-section, indicative of the interaction probability between particles, must be positively defined. As mentioned in Ref.~\cite{Klein:2020jom}, the photon GTMD is not always positive definite. Consequently, it is essential to demonstrate that dilepton production via photon-photon fusion is invariably guaranteed to yield a positive definite cross-section. Based on Eq.~(\ref{eq:cos}), it is interesting to show that 
\begin{eqnarray}
\frac{d\sigma}{dy_{1}dy_{2}d^{2}P_{\perp}d^{2}q_{\perp}d^{2}b_{\perp}} 
& \geq &  \frac{1}{(2\pi)^{2}}\left(\frac{Z^{2}}{\pi^{2}}\right)^{2}\int d^{2}k_{1\perp}d^{2}k_{1\perp}^{\prime}e^{ib_{\perp}\cdot\Delta_\perp}\frac{F_A(k_{1})}{k_{1}^{2}}\frac{F(k_{2})}{k_{2}^{2}}\frac{F_A(k_{1}^{\prime})}{{k^{\prime}}_{1}^{2}}\frac{F_A(k_{2}^{\prime})}{{k^{\prime}}_{2}^{2}}\kaperp\kbperp\kcperp\kdperp\nonumber \\
 &  & \times\frac{4\alpha_{{\rm e}}^{4}}{\hat{s}^{2}}\left[\cos(\phi_{\kaperp}-\phi_{\kbperp}+\phi_{\kcperp}-\phi_{\kdperp})-\cos(\phi_{\kaperp}+\phi_{\kbperp}+\phi_{\kcperp}+\phi_{\kdperp}-4\phi_{P})\right].
\end{eqnarray}
By using the following identities
\begin{eqnarray}
 &  & 2\sin(\phi_{\kaperp}+\phi_{\kcperp}-2\phi_{P})\sin(\phi_{\kbperp}+\phi_{\kdperp}-2\phi_{P})\nonumber \\
 & = & \cos(\phi_{\kaperp}-\phi_{\kbperp}+\phi_{\kcperp}-\phi_{\kdperp})-\cos(\phi_{\kaperp}+\phi_{\kbperp}+\phi_{\kcperp}+\phi_{\kdperp}-4\phi_{P}),
\end{eqnarray}
and 
\begin{equation}
\sin(\phi_{\kaperp}+\phi_{\kcperp}-2\phi_{P})=\sin(\phi_{\kaperp}-\phi_{P})\cos(\phi_{\kcperp}-\phi_{P})+\cos(\phi_{\kaperp}-\phi_{P})\sin(\phi_{\kcperp}-\phi_{P}) \, ,
\end{equation}
one can rewrite the cross section as
\begin{eqnarray}
\frac{d\sigma}{dy_{1}dy_{2}d^{2}P_{\perp}d^{2}q_{\perp}d^{2}b_{\perp}}\geq\left(\frac{Z^{2}}{\pi^{2}}\right)^{2}\frac{8\alpha_{{\rm e}}^{4}}{(2\pi)^{2}\hat{s}^{2}}\left(G^{12}+G^{21}\right)\left(G^{12*}+G^{21*}\right),
\end{eqnarray}
where we define unit vector $\hat{l}_{i}$ which has the
angle $\phi_{l_{i}}=\phi_{k_{i\perp}}-\phi_{P}$ and $G^{ik}$ is defined as $G^{ik}=\int d^{2}\kaperp e^{i\kaperp\cdot b_\perp} F_A(k_1)F_A(k_2) |\kaperp||\kcperp|\hat{l_{1}}^{i}\hat{l_{2}}^{k}$. It is then obvious to see that the cross-section is positive definite, since the right hand side is essentially related to the square of an amplitude.

\subsection{Phenomenological Studies}
\label{sec::phe}
Based on the above unified GTMD factorized framework, we discuss phenomenological studies of dilepton production in this subsection in detail. First, we elaborate on the various dilepton anisotropies and show the numerical predictions. Then, we include the Sudakov effect, which resums the soft photon radiation emitted by the lepton pairs. In addition, we take the neutron emission effect in ultra-peripheral collisions (UPCs) into account. Finally, we discuss the incoherent contribution of the photon Wigner distribution.

\subsubsection{Anisotropies}

\begin{figure}[!ht]
\centering{
    \includegraphics[width=0.8\textwidth]{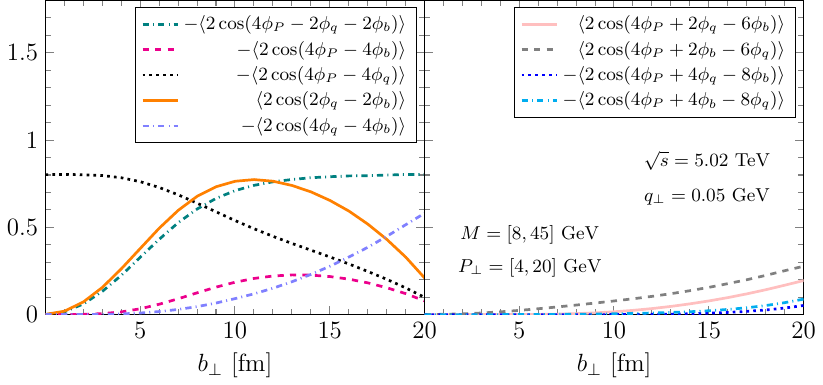}
    }
    \caption{
The prediction of different dilepton anisotropies as a function of $b_\perp$ with pair mass $M=8-45$ GeV, fixed $q_\perp=0.05$ GeV and $P_\perp=4-20$ GeV at the LHC kinematics. The rapidities of the dilepton are integrated over the regions $[-1,1]$. 
    }
    \label{fig:cos_LHC_b}
\end{figure}

\begin{figure}[!ht]
\centering{
    \includegraphics[width=0.8\textwidth]{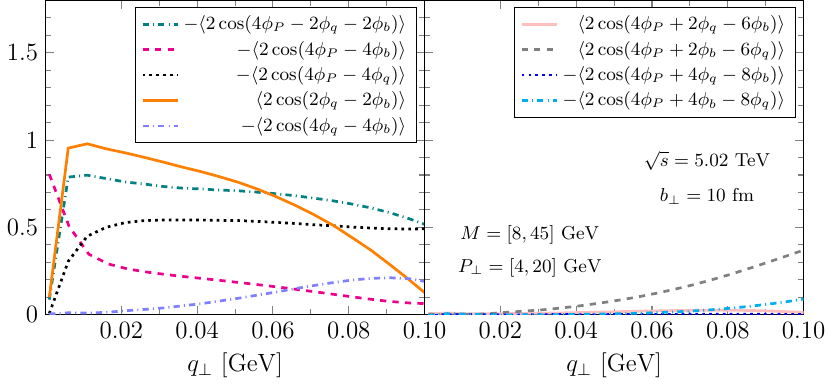}
    }
    \caption{Predictions of different dilepton anisotropies are presented as a function of $q_\perp$ for a pair mass $M=[8,45]$ GeV, with $b_\perp$ fixed at $10$ fm and $P_\perp$ ranging from $4$ to $20$ GeV, within the LHC kinematics. The rapidities of the dileptons are integrated over the interval $[1,1]$.    
      }
    \label{fig:cos_LHC_qt}
\end{figure}

\begin{figure}[!ht]
\centering{
    \includegraphics[width=0.8\textwidth]{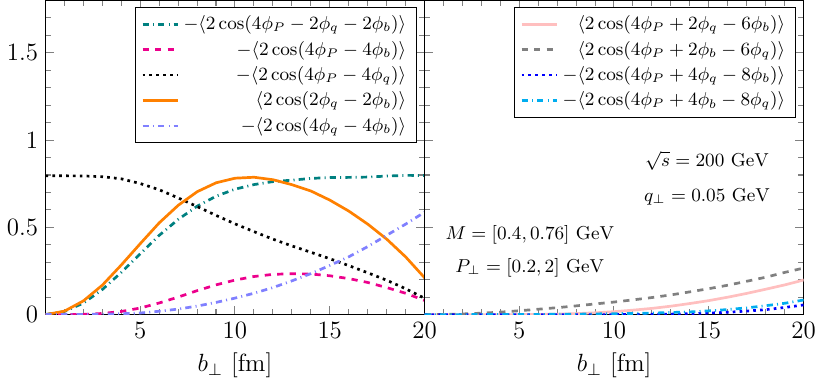}
    }
    \caption{Under RHIC kinematics, the anisotropies of dileptons are predicted as functions of $b_\perp$, with the pair mass $M$ ranging from $0.4$ to $0.76$ GeV, $q_\perp$ fixed at $0.05$ GeV, and $P_\perp$ between $0.2$ and $2$ GeV. The rapidities of the dileptons are integrated over the interval $[-1, 1]$.
     }
    \label{fig:cos_RHIC_b}
\end{figure}

\begin{figure}[!ht]
\centering{
    \includegraphics[width=0.8\textwidth]{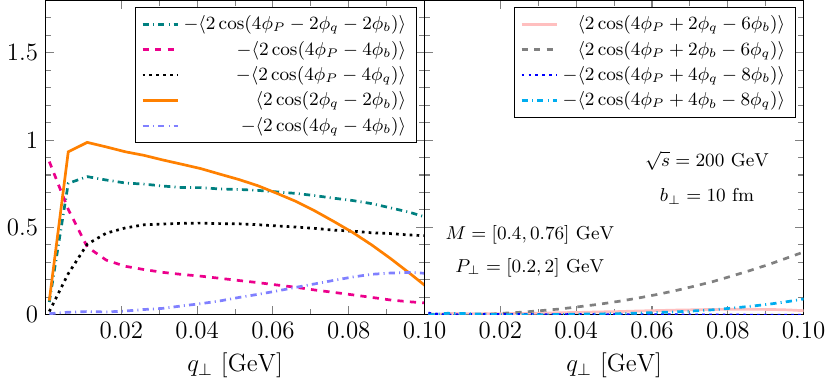}
    }
    \caption{ For RHIC kinematics, the anisotropies of dileptons are predicted as functions of $q_\perp$, with the pair mass $M$ ranging from $0.4$ to $0.76$ GeV, $b_\perp$ fixed at $10$ fm, and $P_\perp$ between $0.2$ and $2$ GeV. The rapidities of the dileptons are integrated over the interval $[-1, 1]$.    
      }
    \label{fig:cos_RHIC_qt}
\end{figure}

In the following, we explore the anisotropy of the dilepton production process $\gamma+\gamma\rightarrow l^+ + l^-$. The cross-section contains various anisotropies. For example, there are a lot of studies about the angular correlation between $P_\perp$ and $b_\perp$, and the angular correlation between $P_\perp$ and $q_\perp$, which results in the commonly measured $\cos(4\phi_P-4\phi_q)$ and $\alpha$ distribution, as shown in Refs.~\cite{Li:2019yzy,Li:2019sin, Xiao:2020ddm, Klein:2020jom,Klein:2020fmr,Klusek-Gawenda:2020eja, Brandenburg:2020ozx, Shao:2022stc, Shao:2023zge}.

Based on the expression of the hard factor $\mathcal{H}^{ijkl}(\vec P_\perp)$ in Eq.~(\ref{hard2}) and its projection properties when contracting with the impact parameter ($\vec b_\perp$) and the transverse momentum of the final lepton pair ($\vec{q}_\perp$), we can project the cross-section in the following general Fourier harmonic series
\begin{eqnarray}
\sigma 
&=& \sigma_0+\sigma_{2qb} \cos (2\phi_q -2\phi_b)  + \sigma_{4qb} \cos (4\phi_q -4\phi_b)  + \sigma_{4qP} \cos (4\phi_q -4\phi_P)    \nonumber   \\
&& +\sigma_{4bP} \cos (4\phi_b -4\phi_P)  + \sigma_{qbP}   \cos (2\phi_q +2\phi_b-4\phi_P) +...,
\end{eqnarray}
where the harmonic coefficients $\sigma_{2qb}, \sigma_{4qb}, ...$ contains the detailed quantum information of the photon flux. From the GTMD factorization framework, we find that linearly polarized photons can generate various anisotropies with considerable magnitudes from the correlation among azimuthal angles of $q_\perp$, $P_\perp$, and $b_\perp$. Not only can we reproduce anisotropies studied in the previous works, but we have also found several other interesting correlations arising from the inner products of $P_\perp$, $q_\perp$ and $b_\perp$. These anisotropies can be divided into two categories. 

The first category of correlation is independent of $P_\perp$. When the $O^{ijkl}$ term from the hard factor $\mathcal{H}^{ijkl}(\vec{P}_\perp)$, an isotropic term associated with $P_\perp$, couples with two incoming Wigner photon distributions, it results in two significant angular correlations: $\cos(2\phi_q - 2\phi_b)$ and $\cos(4\phi_q - 4\phi_b)$. More precisely, when we analyze the cross-section in momentum space, these correlations originate from the contraction between the hard factor and the linearly polarized photon distribution via the phase factor $e^{i\vec{b}_\perp \cdot \vec{\Delta}_\perp}$. This second category involves the $\cos(4\phi_{P})$ harmonics, which comes from the tensor contraction between the $\Omega^{ijkl}(\vec P_\perp)$ term in the hard factor and the incoming linearly polarized photon distributions. Two anisotropies proposed in Refs~\cite{Li:2019yzy, Xiao:2020ddm}, namely $\cos(4\phi_P-4\phi_q)$ and $\cos(4\phi_P-4\phi_b)$, originate from this mechanism and belong to this category. In addition, we also find that the same mechanism can give rise to several novel angular correlation of three final observables $P_\perp$, $q_\perp$, and $b_\perp$, such as $\cos (4\phi_P-2\phi_q -2\phi_b)$, $\cos (4\phi_P-6\phi_q +2\phi_b)$, and $\cos (4\phi_P-6\phi_b +2\phi_q)$, and etc. 

To extract the harmonic coefficients, we can use the following definition and compute the average value of the harmonics $\cos(...)$ 
\begin{equation}
2\langle \cos(...)  \rangle = \frac{2 \int d\mathcal{P.S.}  \cos(...)    \frac{d\sigma}{dy_{1}dy_{2}d^{2}P_{\perp}d^{2}q_{\perp}d^{2}b_{\perp}}}{ \int d\mathcal{P.S.}  \frac{d\sigma}{dy_{1}dy_{2}d^{2}P_{\perp}d^{2}q_{\perp}d^{2}b_{\perp}}},
\end{equation}
where $\int d\mathcal{P.S.}$ represents the phase space integrals. In  computing the angular correlations with given $q_\perp$ and $b_\perp$, we define it as $\int dy_{1}dy_{2}d^{2}P_{\perp}d \phi_q d \phi_b$. We show predictions for the anisotropies as a function of $b_\perp$ with fixed $q_\perp = 50$ MeV at the LHC in the Fig.~\ref{fig:cos_LHC_b}. As shown in these plots, we find that there are nine azimuthal angular correlations, which are numerically significant. As discussed above, the first category includes two angular correlation terms $2\langle \cos(2\phi_q-2\phi_b) \rangle $ and $2\langle \cos(4\phi_q-4\phi_b) \rangle $.  $2\langle \cos(2\phi_q-2\phi_b) \rangle $ has a sizeable positive contribution. As the impact parameter $b_\perp$ increases, it increases in the region $b_\perp<2R_A$, and peaks at $2R_A$, then decreases in the region $b>2R_A$. The $2\langle \cos(4\phi_q-4\phi_b) \rangle $ term is usually negative, and its magnitude increases as the impact parameter $b_\perp$ increases. 

The second category of the correlations are related to the $\cos(4\phi_P)$ harmonics, and it can be further subdivided into three different types. The first type is the angular correlations between $P_\perp$ and $q_\perp$. It includes one angular correlation term, i.e., $2\langle \cos(4\phi_P-4\phi_q) \rangle $. The second type is the angular correlation between $P_\perp$ and $b_\perp$, namely, $2\langle \cos(4\phi_P-4\phi_b) \rangle $. There are lots of related discussions found in Refs.~\cite{Li:2019yzy, Li:2019sin, Xiao:2020ddm, Shao:2022stc, Shao:2023zge} about these two types of correlations. The last type is the angular correlation that mixes all three vectors, $q_\perp$, $b_\perp$, and $P_\perp$. Here, we show five angular correlations as examples. Among them, the $2\langle \cos(4\phi_P-2\phi_q-2\phi_b) \rangle $ term has the largest magnitude. This correlation is negative and its magnitude increases as the impact parameter $b_\perp$ increases, and the value tends to be constant in the region $b>2R_A$. $2\langle \cos(4\phi_P+2\phi_b-6\phi_q) \rangle $ and $2\langle \cos(4\phi_P+2\phi_q-6\phi_b) \rangle $ are positive, while $2\langle \cos(4\phi_P+4\phi_b-8\phi_q) \rangle $ and  $2\langle \cos(4\phi_P+4\phi_q-8\phi_b) \rangle $ are negative. The amplitudes of these four terms all show similar behaviour, and they increases as the impact parameter $b_\perp$ increases. Similarly, as shown in the Fig.~\ref{fig:cos_LHC_qt}, we can find that these anisotropies show similar magnitude as the function of $q_\perp$ at fixed $b=10$ fm in the LHC.

 At a distance of $b=10$ fm, the colliding nuclei undergo peripheral collisions, which also generates a large number of hadrons. Consequently, the reaction plane of the hadronic events can be used to determine the impact parameter vector $\vec{b}_\perp$ in peripheral collisions. By combining the hadronic reaction plane information with the dilepton measurements, it becomes possible to measure various dilepton angular correlations between $b_\perp$ and $q_\perp$, $P_\perp$ in peripheral collisions.

In addition, we show predictions about anisotropies as a function of $b_\perp$ and $q_\perp$ at RHIC energy in Fig.~\ref{fig:cos_RHIC_b} and Fig.~\ref{fig:cos_RHIC_qt}. In these predictions, we determine the values of the pair mass $M$, the relative transverse momentum of the lepton pair $P_\perp$, and the rapidity region based on the relevant kinematic region on RHIC and LHC. Although the kinematics are different at the RHIC and LHC, the longitudinal momentum fractions at different energies are almost identical, and the photon Wigner distributions at different energies are comparable. Therefore, we can see that the predictions show very similar behaviors at the RHIC and LHC. In principle, there are infinite number of correlations in this series. We only plot the nine examples of the angular correlations in the above plots. In the future, it would be interesting to compare our results with RHIC and the LHC measurements and it will help us to understand the property of the photon Wigner distribution. We note that the mass effect has been neglected. If one consider the mass effect, more correlation can be observed. We leave it for future study.

\subsubsection{The Sudakov Effect}\label{sec:sud}

\begin{figure*}[!ht]
\includegraphics[width=.99\linewidth]{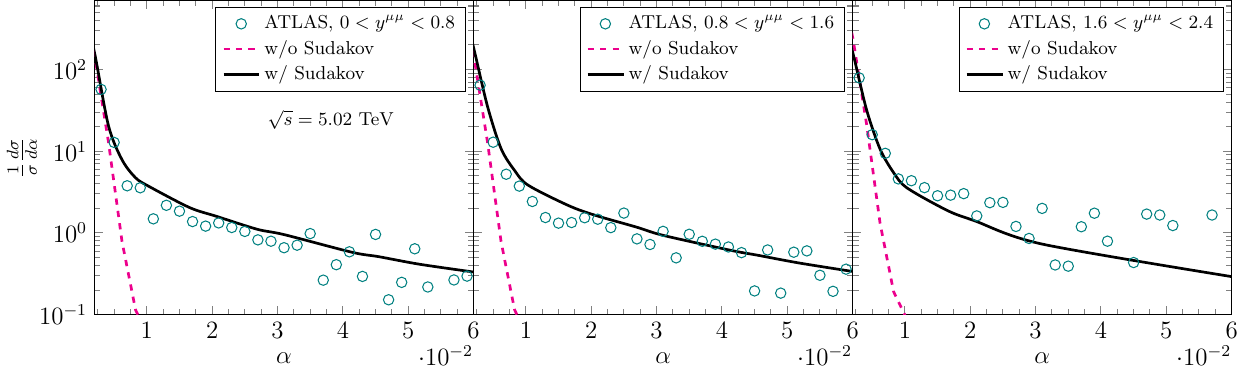}
\caption{Comparisons our results with the ATLAS experimental data~\cite{ATLAS:2016vdy} in the dimuon pair acoplanarity $ \alpha$  distribution in UPCs.}  
\label{fig:atlas_alpha_dis_Sud}
\end{figure*} 

This subsection will discuss the effect of multiple soft-photon radiation, which is also known as the Sudakov soft radiation effect. These soft radiations give rise to large logarithms that must be resummed. The resummation improved cross-section is given as~\cite{Klein:2018fmp, Klein:2020jom, Hatta:2021jcd, Shao:2022stc, Shao:2023zge}
\begin{equation}
\frac{d \sigma}{\pi d \bperp^2 d^2 q_\perp d^2 P_\perp   } 
= \int d^2l_\perp \mathcal S(l_\perp,  Q, m_l) \frac{d \sigma}{\pi d \bperp^2 d ^2q^\prime_\perp  d^2P_\perp  } (q^\prime_\perp=q_\perp -l_\perp).
\end{equation}
As shown in Refs.~\cite{Klein:2020jom, Shao:2023zge}, the Sudakov factor at the leading double logarithm order in the center-of-mass frame of the final lepton pair reads
\begin{equation}
\mathcal S(r_\perp, Q, m_l) = \frac{\alpha_{\rm em}}{2\pi} \left[ 
\ln^2 \frac{Q^2r^2}{c_0^2}  -
\ln^2 \frac{m_l^2r^2}{c_0^2}  \theta \left(m^2_l- r^2/c_0^2 \right)
 \right].
\end{equation}
with the lepton mass $m_l$ and $\alpha_{\rm em}=1/137$. As shown in Ref.~\cite{Klein:2020jom}, the resummation has an analytic form in the momentum space, which can be expressed as follows
\begin{eqnarray}
\mathcal S(l_\perp,  Q, m_l) &=& \int \frac{d^2r_\perp}{(2\pi)^2}e^{il_\perp\cdot r_\perp} e^{-\frac{Q_0^2 r_\perp ^2}{4}} e^{-S_{\alpha}(r_\perp, Q, m_l)} \\
&=&  \frac{ e^{\gamma_0} }{\pi Q_0^2} \left(\frac{c_0^2 Q_0^2}{4Q^2} \right)^{\beta_1} \Gamma(1-\beta_1) ~_1F_1(1-\beta_1, 1, - \frac{l_\perp^2}{Q_0^2}),
\end{eqnarray}
with $\beta_1 = \frac{\alpha_{\rm em}}{\pi} \ln \frac{Q^2}{l_\perp^2+m_l^2} $, $\gamma_0= \frac{\alpha_{\rm em}}{2\pi} \ln^2 \frac{Q^2}{m_l^2}$, $c_0=2e^{-\gamma_E}$, the Euler constant $\gamma_E$, and Hypergeometric function $~_1F_1$. To suppress the large $r_\perp$ contribution and expedite the numerical calculations, we use a Gaussian factor with a Gaussian width $Q_0=10$ MeV. As a consistent check, we find that our numerical results is insensitive to the choice of the width as long as $Q_0$ is kept small.  

In Fig.~\ref{fig:atlas_alpha_dis_Sud}, we present our calculations with and without the Sudakov effect compared to the ATLAS data~\cite{ATLAS:2016vdy}. The Leading Order (LO) calculation could provide a good description of the lepton pair acoplanarity in the small $\alpha$ region and small-$q_\perp$ broadening. However, the cross-sections without the Sudakov factor fall too rapidly at large acoplanarity regions. When considering leptons with larger acoplanarity or larger $q_\perp$, one must include the multiple emissions of soft photons. The calculations that include the Sudakov effect better fit the experimental data at large acoplanarity regions. It demonstrates that Sudakov resummation is critical for accurately computing large acoplanarity regions in dilepton production.

\subsubsection{Dilepton Events with Neutrons Emission}

In this section, we focus on the measurements of dilepton production with neutron emissions and explore the resulting bias in the impact parameter and its corresponding connection with the incoherent photon contribution. Recently, there have been many measurements on dilepton production with neutron tagging from the STAR, CMS, and ATLAS Collaborations in UPCs~\cite{STAR:2019wlg, ATLAS:2020epq, CMS:2020skx, ATLAS:2022srr}. When an event involves neutron emission, these neutrons, which are almost collinear with the beam direction, can be detected by the Zero Degree Calorimeter (ZDC).

Generally, the number of emitted neutrons depends on the impact parameter $b_\perp$ between two heavy ions. As assumed in Ref.~\cite{Baltz:2002pp}, the probability of having $N$ neutron emissions follows the Poisson distribution
\begin{equation}
P(N, b_\perp) = \frac{1}{N!}[P_s(b_\perp)]^N \exp[-P_s(b_\perp)],
\end{equation}
where $P_s(b_\perp)= S/b_\perp^2$ is the lowest-order probability for an excited nucleus to emit one neutron, and $P(N, b_\perp)$ represents the probability of emitting $N$ neutrons. Here, one uses the parameter $S$ to indicate the strength of neutron emissions. As shown in Ref.~\cite{Baur:1998ay}, the Giant-Dipole Resonance model gives an expression for $S$ as follows
\begin{equation}
S=5.45\times 10^{-5} \frac{Z^3(A-Z)}{A^{2/3}}  \text{ fm}^2,
\label{eq:S}
\end{equation}  
with $A =$ atomic mass number and $Z$ being the proton number. For Pb-Pb collisions, the above formula gives $S=108 \text{ fm}^2 $. In addition, one may also choose another parametrization $S=303 \text{ fm}^2 $ which includes all additional radiation mechanism for Pb-Pb collisions~\cite{Baltz:1997di}. These two different prescriptions show different dependency relations between the probability of neutron emission and the impact parameter. They will be referred to as PAR I and PAR II in the subsequent numerical studies, respectively. We find that some observables are sensitive to the integration range of the impact parameter, and we will discuss in detail the effects of these two prescriptions in the next section. Based on the Poisson distribution, we can write the probability of different numbers of neutron emission as follows
\begin{eqnarray}
P_0(b) &=& \exp[-P_s(b)], \\
P_1(b)  &=&P_s(b) \exp[-P_s(b) ],\\
P_M(b) &=& 1- \exp[-P_s(b)],\\
P_X(b) &=& 1- \exp[-P_s(b)] - P_s(b) \exp[-P_s(b)],
\end{eqnarray}	
where $P_0(b)$ is the probability of no neutron emission, $P_1(b)$ is the probability of one neutron emission, $P_M(b)$ is the probability of at least one neutron emission, and $P_X(b)$ is the probability of at least two neutron emissions.

Furthermore, we can categorize the UPC event into different scenarios according to the number of collinearly emitted neutrons detected in the ZDC on both ends. For example, the scenario 
$1nXn$ means that exactly one neutron is detected on one ZDC, and at least two neutrons are detected in the other ZDC. $P_{nn}(b)$ is then defined as the combined probability of various type of neutron emissions from both nuclei as a function of impact parameter. We summarize various scenarios in the following table and plot the probability distributions in the right part of Fig.~\ref{table:1}. As shown in the plot of the combined probabilities, the neutron emission multiplicity biases toward a small value of the impact parameter. This means that the distance between two colliding nuclei is, on average, smaller in the events with higher neutron multiplicity. Therefore, neutron emissions can be used as a tool to select events with different impact parameters.

\begin{figure}[!ht]
    \centering
   \begin{minipage}{0.59\textwidth} 
        \centering
\begin{tabular}{ |c|c|c| } 
\hline
  & Neutrons emission   &  $P_{nn}(b)$\\
  \hline
     1 &  $0n0n$     & $ P_{00}(b)=[P_0(b)]^2$ \\ 
  \hline
       2 & $0n1n$      & $P_{01}(b)=2P_0(b)P_1(b)$ \\ 
  \hline
      3& $0nMn$      & $P_{0M}(b)=2P_0(b)P_M(b)$ \\ 
  \hline
      4& $0nXn$      & $P_{0X}(b)=2P_0(b)P_X(b)$ \\ 
  \hline
     5 &  $1n1n$       & $P_{11}(b)=[P_1(b)]^2$ \\ 
  \hline
      6& $1nXn$      & $P_{1X}(b)=2P_1(b)P_X(b)$ \\ 
  \hline
 7 &  $MnMn$       & $P_{MM}(b)=[P_M(b)]^2$ \\ 
  \hline
 8 &  $XnXn$       & $P_{XX}(b)=[P_X(b)]^2$ \\ 
  \hline
\end{tabular}
    \end{minipage}
        \begin{minipage}{0.35\textwidth} 
        \centering
        \includegraphics[width=\textwidth]{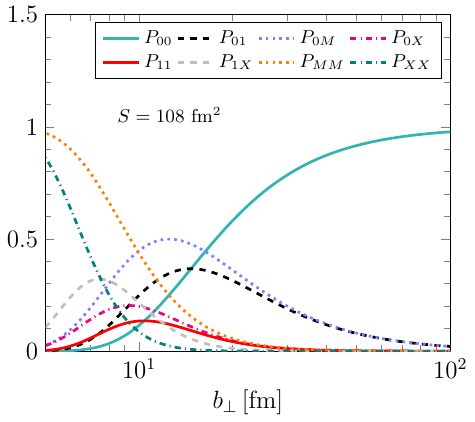} 
          \end{minipage}\hfill
\caption{Left: Table of the combined probability for various neutron emissions. Right: The plot of the combined probabilities as functions of $b_\perp$. }
\label{table:1}
\end{figure} 
In addition, to include the effect of neutron emission in our framework, we cast the cross-section into
\begin{eqnarray}
\frac{d \sigma_{nn}}{d^2P_\perp d^2q_\perp   } 
&=&
\int^\infty_{b_{\min}}  d b_\perp b_\perp \int d\phi_b  \frac{d \sigma}{ d^2 b_\perp d^2P_\perp d^2q_\perp  }  P_{nn}(b_\perp),
\label{eq:ne}
\end{eqnarray}
where $b_{\rm min}$ is defined as the minimum impact parameter value to distinguish between peripheral collisions and the UPCs.
For the case of $0n0n$, we note that the neutron emission probability $P_{00}$ is non-vanishing at large $b_\perp$, resulting in slow numerical convergence when combined with the phase factor. Since the probability of $P_{00}(b_\perp)$ approaches one when the impact parameter is much larger than $R_A$, we can use the following technique to calculate the $0n0n$ case numerically. We first separate the cross-section into two terms as follows
\begin{eqnarray}
\frac{d \sigma_{00}}{d^2P_\perp d^2q_\perp   } &=&\int ^\infty _{b_{\rm min}} db_\perp b_\perp \int d\phi_b   \frac{d\sigma}{d^2b_\perp d^2P_\perp d^2q_\perp } P_{00}(b_\perp) \nonumber \\
&=&\int ^\infty _{b_{\rm min}} db_\perp b_\perp \int d\phi_b \frac{d\sigma}{d^2b_\perp d^2P_\perp d^2q_\perp } - \int ^\infty _{b_{\rm min}} db_\perp b_\perp \int d\phi_b \frac{d\sigma}{d^2b_\perp d^2P_\perp d^2q_\perp }\left[1- P_{00}(b_\perp) \right],
\end{eqnarray}
where the second term is convergent rapidly in numerical computation, and the first term is the full UPC contribution
\begin{equation}
 \frac{d\sigma^{\rm UPC}}{d^2P_\perp d^2q_\perp }= \int ^\infty _{b_{\rm min}} db_\perp b_\perp \int d\phi_b \frac{d\sigma}{d^2b_\perp d^2P_\perp d^2q_\perp }.
\end{equation}
We can analytically perform the integration over impact parameter from $b_{\rm \min}$ to infinity for $\sigma^{\rm UPC}$ by following the same technique as suggested in Ref.~\cite{Klein:2020jom}.

\subsubsection{The Incoherent Contribution to the Photon Distribution}\label{sec:incoh}

\begin{figure}[!ht]
    \includegraphics[width=0.49\textwidth]{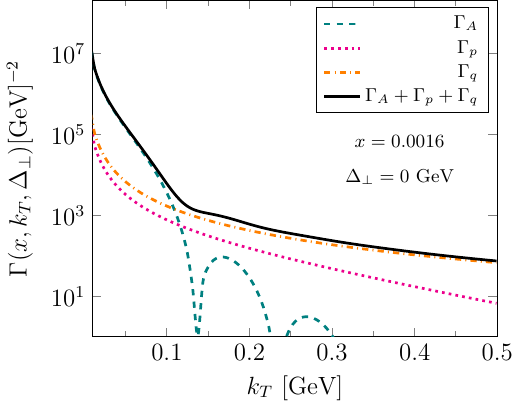}
    \caption{The illustration of the coherent and incoherent contributions to the GTMD photon distribution, with the coherent nucleus represented by dashed lines, the incoherent proton contribution by dotted lines, and the incoherent quark contribution by dash-dotted lines. The solid lines depict the combined distribution of coherent and incoherent contributions. We set the momentum fraction at $x=0.0016$ and $\Delta_\perp$ to $0$ GeV.}
    \label{fig:GTM}
\end{figure}

Let us discuss the incoherent contributions to the photon distribution and its connection to the neutron emission events. For dilepton production from the photon-photon fusion process, when the momentum imbalance $q_\perp$ of the lepton pair is small compared to $1/R_A$ (where $R_A$ is the radius of the nucleus), the nucleus, as a whole, can be treated as the source of photons. This contribution is commonly referred to as the coherent contribution. However, as the momentum imbalance of the lepton pair increases, the transverse momenta carried by the incoming photons should also increase accordingly. Therefore, according to the uncertainty principle, the photon with large transverse momentum no longer originates solely from the coherent nuclear radiation process but can be emitted from the inner structure of the nucleus, such as the protons or the quarks inside the protons and neutrons. These contributions are known as the incoherent contributions.

If quasi-photons with relatively large transverse momenta are radiated from the interior of heavy nuclei, these emissions can likely lead to the disintegration of these nuclei. Consequently, the structural integrity of the heavy nuclei can serve as a key indicator in determining whether this incoherent interaction occurs or not. As discussed in the above section, lots of experimental data with the neutron emission effect have been observed by STAR, CMS, and ATLAS Collaborations~\cite{STAR:2019wlg, ATLAS:2020epq, CMS:2020skx, ATLAS:2022srr}. For the $0n0n$ event, the outgoing nuclei are assumed to still remain intact, thus the dominant contribution is from the coherent photon emissions. In the $MnMn$ and $0nMn$ events in which ZDC can detect a number of neutrons, the outgoing nuclei are disintegrated, and incoherent photon interactions should take place. Therefore, we should include the incoherent contribution in our framework to provide a complete description of the experimental data with neutron emissions.

To account for the incoherent contributions, it is still reasonable to treat the incoming photon as a quasi-real particle and employ the generalized equivalent photon approximation for the calculation. The only difference is that we are treating the nucleus as an incoherent collection of electric charges. Consequently, considering the coherent and incoherent contributions together,  one can modify the corresponding $\Gamma^{ij}(x,k_{T},\Delta_\perp)$, and write it as
\begin{equation}
\Gamma^{ij}(x, \vec k_{T},\vec \Delta_\perp)= \Gamma_A^{ij}(x, \vec k_{T},\vec \Delta_\perp)+\Gamma_p^{ij}(x, \vec k_{T},\vec \Delta_\perp)+\Gamma_q^{ij}(x, \vec k_{T},\vec \Delta_\perp),
\end{equation}
where the first term is the coherent contribution from the nucleus. The second and last terms are the incoherent contributions from the protons and quarks inside nucleons, respectively.  When the protons inside the heavy nucleus are viewed as the photon sources, the corresponding total GTMD of the nucleus is given as the incoherent sum of the photon distributions from individual protons
\begin{equation}
\Gamma_p^{ij}(x,\vec k_{T},\vec \Delta_\perp)= Z\frac{\alpha_{\rm em}}{\pi^2} k^i_{\perp}{k^\prime}^j_{\perp}  \frac{F_p(k)}{k^2} \frac{F_p(k^{\prime})}{{k^{\prime}}^{2}},
\label{eq:xfp}
\end{equation}
where $F_p(k)$ is the proton form factor.
When the photon is radiated incoherently by the quark inside protons or neutrons, we can write the photon GTMD by treating the quarks as point particles, which is presented as follows
\begin{equation}
\Gamma_q^{ij} (x,\vec k_{T},\vec \Delta_\perp) = 
\left[Z \sum e_{q/p}^2+ (A-Z) \sum e_{q/n}^2 \right]
\frac{\alpha_{\rm em} }{\pi^2} k^i_{\perp}{k^\prime}^j_{\perp}
 \frac{1}{k^{2}} \frac{1}{{k^{\prime}}^{2}},
\end{equation}
where $e_q$ is the charge number of quark from protons and neutrons.  In the above equation, the first and second terms correspond to the contribution of the quark inside protons and neutrons, respectively. According to the experimental cuts used in the measurement, the transverse momentum imbalance of the final lepton pairs $q_\perp$ is expected to be less than $1.2$ GeV, which means the quantum fluctuations within nucleons are comparatively small and the QCD evolution effect may be neglected. Therefore, we can use the valence quark model to simplify the calculation. In this model, a proton comprises two up quarks and one down quark, while a neutron consists of one up quark and two down quarks, with these quarks inside nucleons taking the same momenta. 

In our numerical calculations, we employ the following parametrizations of the nuclear form factor~\cite{Davies:1976zzb} 
$F_A(k)=\frac{3}{(kR_A)^3}\frac{1}{1+a^2k^2}\left[\sin(kR_A)-kR_A\cos(kR_A)\right]$, where $a=0.71$ fm, and $R=6.6$ fm for $Pb$ and $R=6.4$ fm for $Au$ and the proton form factor~\cite{Andivahis:1994rq,Klein:2003vd} 
$F_p(k) =\frac{1}{[(k^2/0.71 \text{ GeV}^2)+1]^2}$.

For comparison, we plot three different contributions (coherent, incoherent proton part, and incoherent quark part) to the GTMD photon distribution in Fig.~\ref{fig:GTM}. The momentum fraction $x$ in this plot is estimated with $p_{1\perp}=p_{2\perp}=4$ GeV and $\Delta_\perp=0 \text{ GeV}$ at midrapidity $y_1=y_2=0$ at LHC energy. We observe that the photon distribution originating from the coherent nucleus $\Gamma_A$ dominates at low $k_T$ values, while the incoherent contributions are a couple of magnitudes smaller than the coherent nucleus contribution. In the large $k_T$ regime, the coherent contribution $\Gamma_A$ falls off rather rapidly as $k_T$ increases, and the incoherent distribution becomes much larger than the coherent contribution when $k_T > 0.15$ GeV. Furthermore, the quark contribution to the photon distribution $\Gamma_q$ is larger than that from protons $\Gamma_p$. For the sum of coherent and incoherent contributions, in the small $k_\perp$ regions, the nucleus acts as the dominant photon source, which means the modified GTMD is consistent with the traditional GTMD in the small $k_\perp$ region. In the large $k_\perp$ region, the incoherent distribution from quarks takes over. It is interesting to note that the combined photon GTMD distribution is proportional to $Z^2$ in the small transverse momentum region and proportional to $(A-Z) \sum e_{q/n}^2+Z \sum e_{q/p}^2=\frac{2}{3}(A-Z) + Z$ in the asymptotically large transverse momentum region. The incoherent emission of photons is rare compared to the coherent emission, yet it is more efficient in generating high $k_T$ photons than the coherent emission.

For dilepton production, if the transverse momentum imbalance of the final lepton pairs $q_\perp$ is small, the two colliding initial photons typically possess low transverse momenta, which in turn suggests that the photons originate predominantly from the coherent nucleus. Under such conditions, the photons ``see" the nucleus as a whole. Conversely, when $q_\perp$ is large, either or both of the initial photons are likely to have high transverse momenta and are typically emitted by protons or quarks. This means that these high-momentum photons can allow us to perceive the internal structures within the nucleus. Therefore, this incoherent process is likely to excite the nucleus, causing it to emit one or more neutrons in the beam direction, which can subsequently be recorded by the ZDC detector. The incoherent contribution is expected to be the dominant production mechanism for $q_\perp$ broadening in dilepton production for events with one or more neutron emissions, such as $0nMn$ and $MnMn$. In the next section, we will compare our numerical calculations with experimental data involving neutron tagging and show that the inclusion of the incoherent contribution is required if one wishes to describe both the data with and without neutron emissions.

\subsection{Comparisons with the RHIC and LHC data}
\label{sec:com}

In this section, we compare our numerical results with the dilepton experimental data collected in a wide variety of measurements. Firstly, we provide a numerical comparison with the acoplanarity and $q_\perp$ broadening results. 
Furthermore, we compare our results with the neutron-tagging data from STAR, CMS, and ATLAS Collaborations. Also, we turn on the incoherent contributions to account for enhancements in the large $\alpha$ region when neutron emissions are detected. To minimize various uncertainty, we choose to normalize both the computed cross-section and the experimental data to unity. In addition, as the measurements conducted thus far have focused solely on the angular correlation at a fixed impact parameter $b_\perp$ without specifying its orientation, we integrate over the azimuthal angle of $b_\perp$ in the subsequent numerical calculations when comparing with the data.

\subsubsection{Acoplanarity and Transverse Momentum Broadening}\label{sec:aco}

\begin{figure*}[!ht]
\includegraphics[width=1\linewidth]{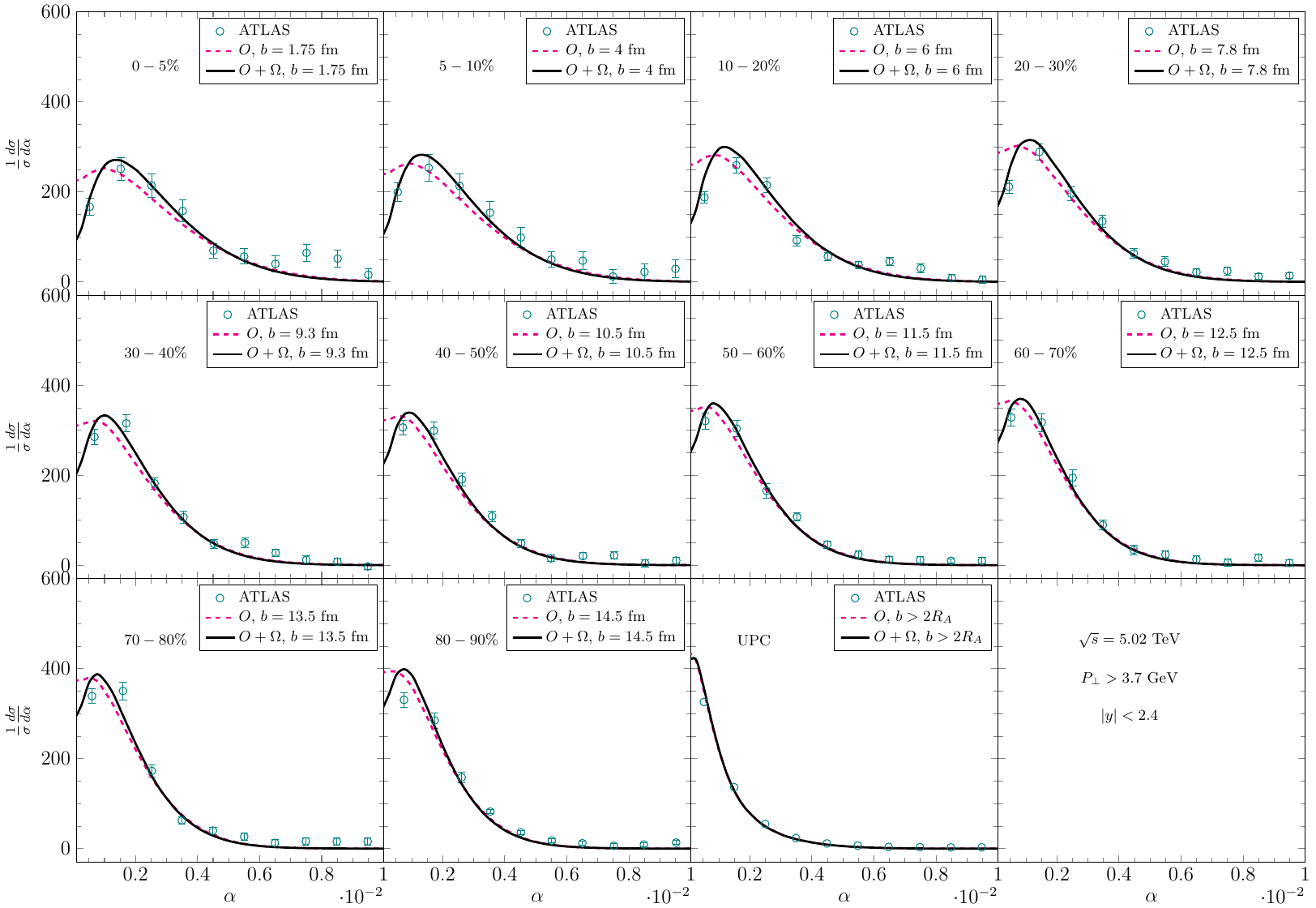}
\caption{Comparisons of our results with the ATLAS experimental data~\cite{ATLAS:2022yad} in the dimuon pair acoplanarity $\alpha$ distribution from central collisions to UPC.}  
\label{fig:atlas_alpha}
\end{figure*}

\begin{figure*}[!ht]
\includegraphics[width=1\linewidth]{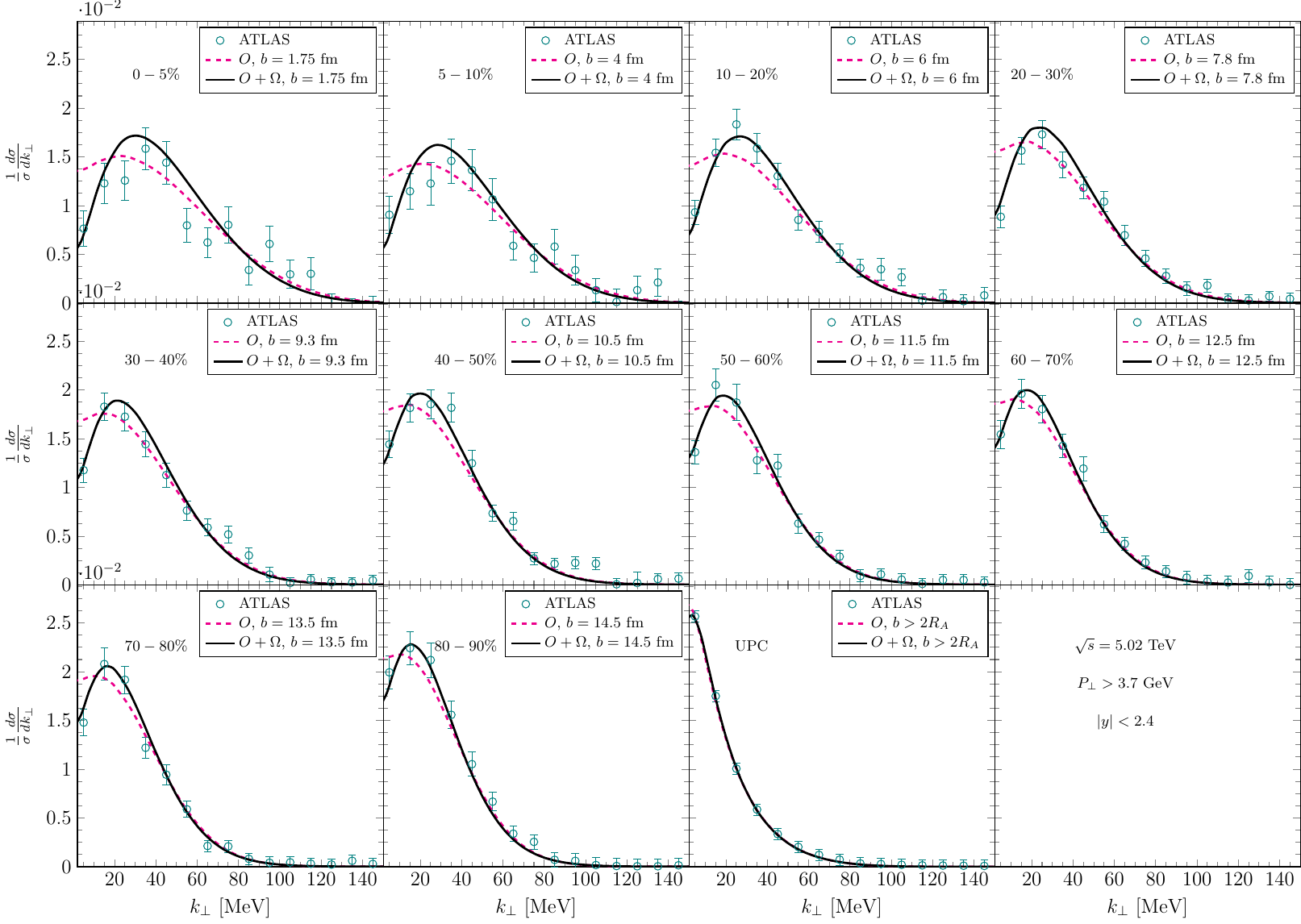}
\caption{Comparisons of our results with the ATLAS experimental data~\cite{ATLAS:2022yad} in the dimuon pair $k_\perp$ distribution from central collisions to UPC.}  
\label{fig:atlas_kt}
\end{figure*}

\begin{figure*}[!ht]
\includegraphics[width=.45\linewidth]{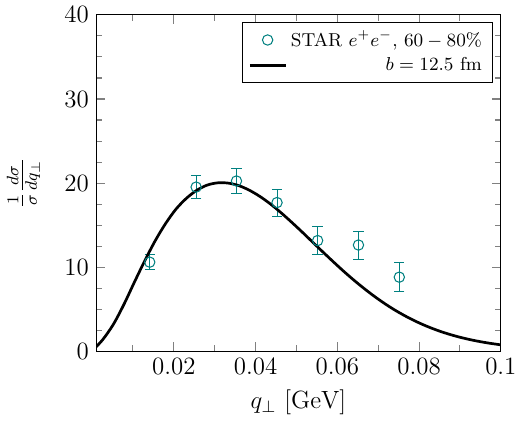}
\caption{Comparisons of our results with the STAR experimental data~\cite{STAR:2019wlg} in the dilepton pair transverse momentum imbalance  $q_\perp$ distribution at $60-80\%$ central collisions. }  
\label{fig:star_qt}
\end{figure*} 

In this subsection, we provide quantitative analyses on the $\alpha$ distribution and its related observable $k_\perp$ in comparison with the ATLAS experimental data. We then further calculate the $q_\perp$ distribution and compare it with the STAR measurements for studying the dilepton broadening effects.

The acoplanarity $\alpha$ is the physical observable describing the azimuthal angular correlation between the lepton pairs. It is defined as 
\begin{equation}
\alpha = 1- |\Delta \phi |/\pi,
\end{equation}
where $\Delta\phi=\phi_1-\phi_2$ is the angle difference between the azimuthal angle of lepton and anti-lepton $\phi_1$ and  $\phi_2$. In Fig.~\ref{fig:atlas_alpha}, we compare our results with the ATLAS experimental data~\cite{ATLAS:2022yad} in the dimuon pair acoplanarity $\alpha$ from central collisions to UPC. The values of the impact parameter corresponding to the different centralities are extracted using the Glauber model. Since the normalized central collision results are almost identical for the impact parameters $b_\perp=0$, $1.75$, $3.5$ fm, thus we simply use the average $b_\perp=1.75$ fm for the $0-5\%$ centrality bin. One observes that our results can perfectly describe the experimental data. Although the dominant contribution is attributed to the angle average of $\phi_P$ term $O^{ijkl}$, it does not capture the dip observed in very small $\alpha$ regions at central collisions. By including the angular correlation $\cos 4\phi_P$ term $\Omega^{ijkl}$, we can describe the dip across various centrality ranges. Hence, the role of the $\Omega^{ijkl}$ term is crucial in accurately describing the dip in central collisions. As the impact parameter $b_\perp$ increases, the significance of the $\Omega^{ijkl}$ term diminishes. Notably, in the case of UPC with large impact parameters, the dominant contribution stems from the $O^{ijkl}$ term, rendering the $\Omega^{ijkl}$ term negligible. In addition, we can find that the peak of $\alpha$ distribution is shifted to smaller values of $\alpha$ from central collisions to UPC. We note that these numerical findings can also be derived from the computations in Refs.~\cite{Zha:2018tlq, Li:2019yzy, Li:2019sin, Wang:2021kxm, Wang:2022gkd, Shao:2023zge}, with this dip arising from the presence of the $\cos 4\Delta \phi$ term.

Besides the $\alpha$ distribution, ATLAS experiments have also used another physical observable $k_\perp$ which is closely related to $\alpha$. This observable can provide us a good estimate of the momentum imbalance and it is defined as 
\begin{equation}
k_\perp=\pi \alpha P_{\perp},
\end{equation}
where $P_\perp$ is the relative transverse momentum of the lepton pair. As shown in Fig.~\ref{fig:atlas_kt}, we compare our results with the ATLAS experimental data~\cite{ATLAS:2022yad} in the dimuon pair acoplanarity $k_\perp$, ranging from central collisions to UPC. Our results offer a good description of the data across a wide range. We observe that the $k_\perp$ distribution shows similar behaviours with the $\alpha$ distribution,  where the dip at small $k_\perp$ for small $b_\perp$ can be explained by including the contribution of the anisotropic $\Omega^{ijkl}$ term.

Additionally, we present our calculation alongside the STAR experimental data~\cite{STAR:2019wlg} as shown in Fig. \ref{fig:star_qt}, in the transverse momentum imbalance $q_\perp$ distribution of dilepton pairs at $60-80\%$ central collisions. Our results provide a satisfactory description of the data. Given that the momentum imbalance $q_\perp$ is an angle-independent observation, only the average term $O^{ijkl}$ of the hard factor contributes, and the $\Omega^{ijkl}$ term of the hard factor vanishes.

Therefore, the above figures show that the dilepton acoplanarity and the broadening of the momentum imbalance $q_\perp$ encode the initial photons' transverse momentum information and they allow us to further explore the $5$-dimensional photon Wigner distribution function and GTMD.

\subsubsection{The Neutron Emission effect in the Small $\alpha$ Region}

\begin{figure*}[!ht]
\includegraphics[width=.46\linewidth]{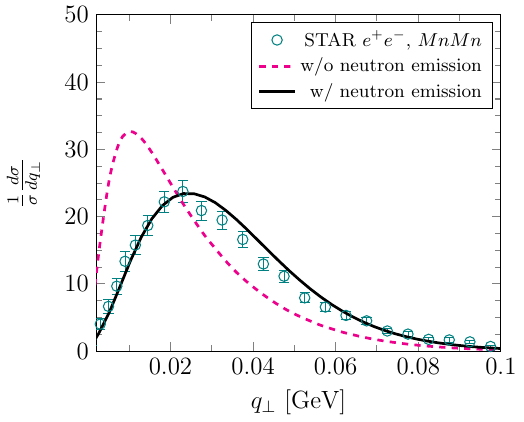}
\includegraphics[width=.46\linewidth]{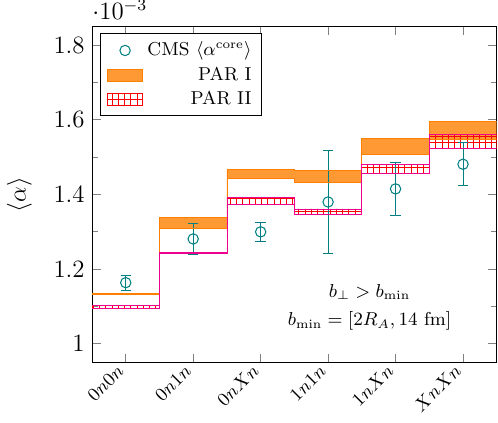} 
\caption{Left: Comparison of our numerical results for the transverse momentum imbalance $q_\perp$ distribution of dielectron pairs, with and without triggering the neutron emission, with the STAR experimental data~\cite{STAR:2019wlg} measured in UPCs with neutron emissions; Right: Comparisons of our results for the dimuon pair $\langle \alpha \rangle$ with the CMS experimental data~\cite{CMS:2020skx}.}  
\label{fig:star_upc}
\end{figure*}

\begin{figure*}[!ht]
\includegraphics[width=.92\linewidth]{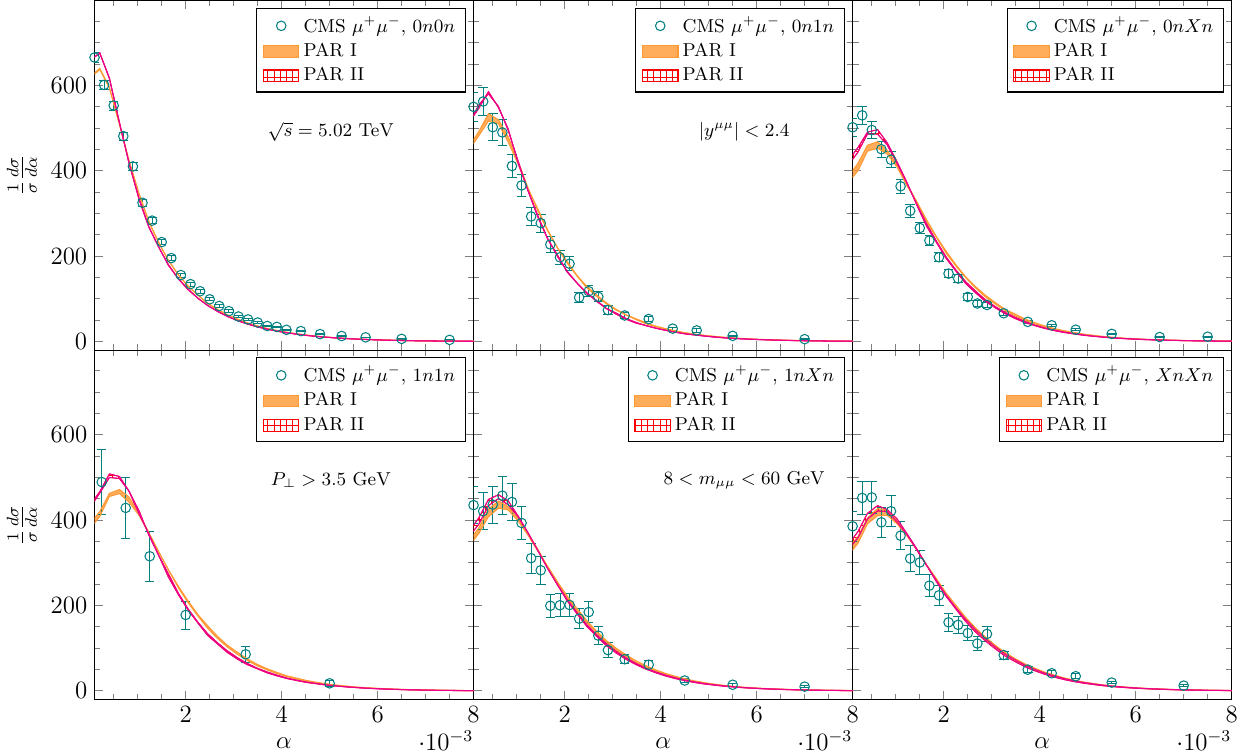}
\caption{Comparisons of our results with the CMS experimental data~\cite{CMS:2020skx} for the dimuon pair acoplanarity $ \alpha$ distribution with the neutron emission in UPCs.}  
\label{fig:cms_alpha_dis}
\end{figure*} 
 
In the small $\alpha$ region or low transverse momentum imbalance regime, the neutron multiplicity in UPCs can serve as an indicator of the impact parameter, as more neutron emission biases toward smaller impact parameter. Firstly, we show the impact of neutron emission on the $q_\perp$ distribution at RHIC energy. Secondly, we study the average acoplanarity $\langle\alpha\rangle$ distribution measured at CMS Collaboration. The comparisons of our numerical calculations with the acoplanarity $\alpha$ distribution from CMS Collaboration are also presented for different configurations of neutron emission tags. Lastly, we compare our numerical results with the $\alpha$ distribution measured at ATLAS Collaboration in the large $\alpha$ region.

In the left plot of Fig.~\ref{fig:star_upc}, we compare our numerical results with the neutron-tagged UPC data from the STAR Collaboration~\cite{STAR:2019wlg}. The STAR experiment uses $Mn$ to denote the emission of one or more neutrons. Here, we present two numerical calculations, one including the corresponding probability to account for the neutron emission effect, and the other without. For this calculation, we adopt the prescription given in Eq.~(\ref{eq:S}) (PAR I) and choose $b_{\rm min}=2 R_A$. The other parametrization gives similar results. Since the neutron emission probability function for $MnMn$ is biased towards smaller impact parameter $b_\perp$ values in UPC, we can see from the result in the previous subsection that $q_\perp$ broadening effect is stronger in smaller values of $b_\perp$. Therefore, we observe that including the neutron emission probability shifts the peak of the $q_\perp$ distribution to a higher $q_\perp$ value. We can find that the calculation combined with the neutron emission probability offers an accurate description of the neutron-tagged UPC experimental data from the STAR Collaboration. This calculation shows that the neutron emission probability is important for describing the experimental data with neutron tagging in the low $q_\perp$ regime.

In the right plot of Fig.~\ref{fig:star_upc}, we present the comparison of the average normalized acoplanarity, $\langle \alpha \rangle$, with the CMS data reported in six neutron multiplicity classes. For this calculation, we utilize both parametrizations for neutron emission probabilities: PAR I with $S=108 \text{ fm}^2$ in Eq.~(\ref{eq:S}) for lead ions, and PAR II with $S=303 \text{ fm}^2$. These two prescriptions yield different probabilities of neutron emission scenarios as a function of the impact parameter $b_\perp$. As shown in the figure, these two prescriptions show the right trend and magnitudes comparing to the CMS data. Furthermore, the average acoplanarity value increases as neutron scenarios transition from $0n0n$ to $XnXn$. We observe that the $\langle \alpha \rangle$ depicts a strong dependence on the parameter $S$ in different neutron scenarios, and the average value  $\langle \alpha \rangle$ decreases as the parameter $S$ increases. For example, for the $XnXn$, the peak of the neutron emission probability $P_{nn}(b_\perp)$ shifts towards a smaller impact parameter $b_\perp$ as the parameter $S$ decreases. Furthermore, the edges of the two bands are calculated by varying $b_{\min}$ from $2R_A$ to $14$ fm in Eq.~(\ref{eq:ne}), and we notice that the average value  $\langle \alpha \rangle$ decreases as $b_{\min} $ increases. The average values $\langle \alpha \rangle$ are sensitive to the lower limit of integration $b_{\min}$, especially for high neutron multiplicity classes.

In Fig.~\ref{fig:cms_alpha_dis}, we present the comparison of the normalized $\alpha$ distribution with the CMS data for six neutron multiplicity classes. In this calculation, we employ two different parametrizations for the neutron emission probability: PAR I and PAR II, as previously defined. Both prescriptions can provide a good description of the CMS data. Interestingly, the peak value of the $\alpha$ distribution shifts to larger value from the $0n0n$ class to the $XnXn$ class. In the large $\alpha$ region, we find that these distributions are insensitive to the choice of the parameter $S$. In summary, our analysis reveals that the neutron emission probability is essential in describing the measurement with the neutron tagging.  


\subsubsection{Enhancements from the Incoherent Contributions in the Large $\alpha$ Region}

\begin{figure*}[!ht]
\includegraphics[width=.46\linewidth]{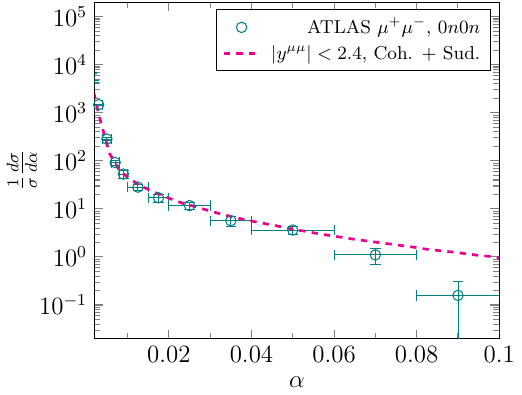}
\caption{Comparisons of our result with the ATLAS experimental data~\cite{ATLAS:2020epq} in the dimuon pair acoplanarity $\alpha$ distribution for $0n0n$ with the rapidities regions of the dimuon $|y^{\mu\mu}|<2.4$.}  
\label{fig:atlas_nm24}
\end{figure*} 

\begin{figure*}[!ht]
\includegraphics[width=.92\linewidth]{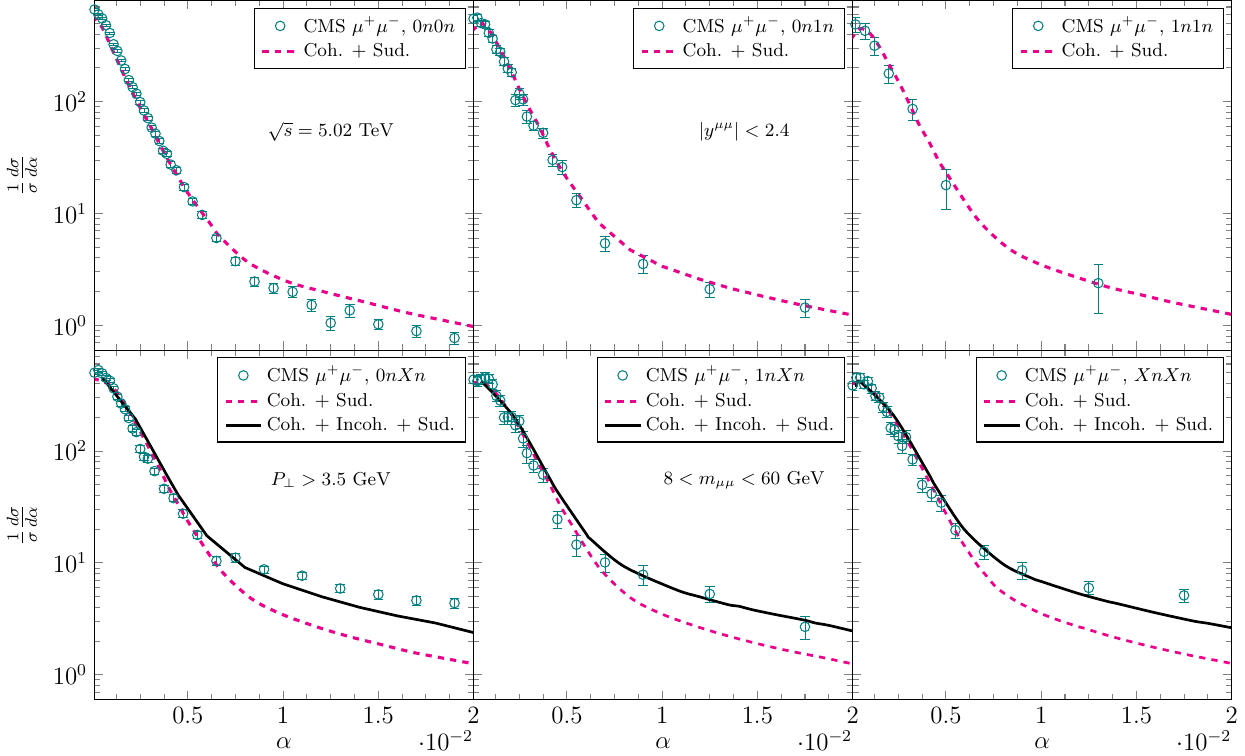}
\caption{Comparison of our results with the CMS experimental data~\cite{CMS:2020skx} in the dimuon pair acoplanarity $\alpha$ distribution with neutron emissions in UPCs.}  
\label{fig:cms_alpha_dis_Sud}
\end{figure*}

\begin{figure*}[!ht]
\includegraphics[width=0.9\linewidth]{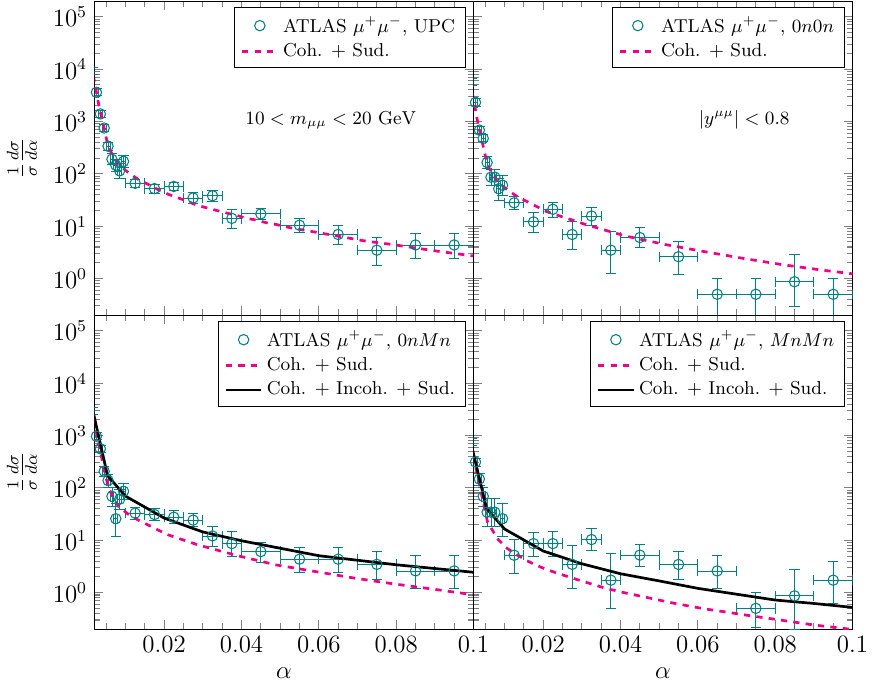}
\caption{Comparison of our results with the ATLAS experimental data~\cite{ATLAS:2020epq} in the dimuon pair acoplanarity $\alpha$ distribution, taking into account neutron emission in UPCs.}  
\label{fig:atlas_mumu_nm08}
\end{figure*}

\begin{figure*}[!ht]
\includegraphics[width=0.9\linewidth]{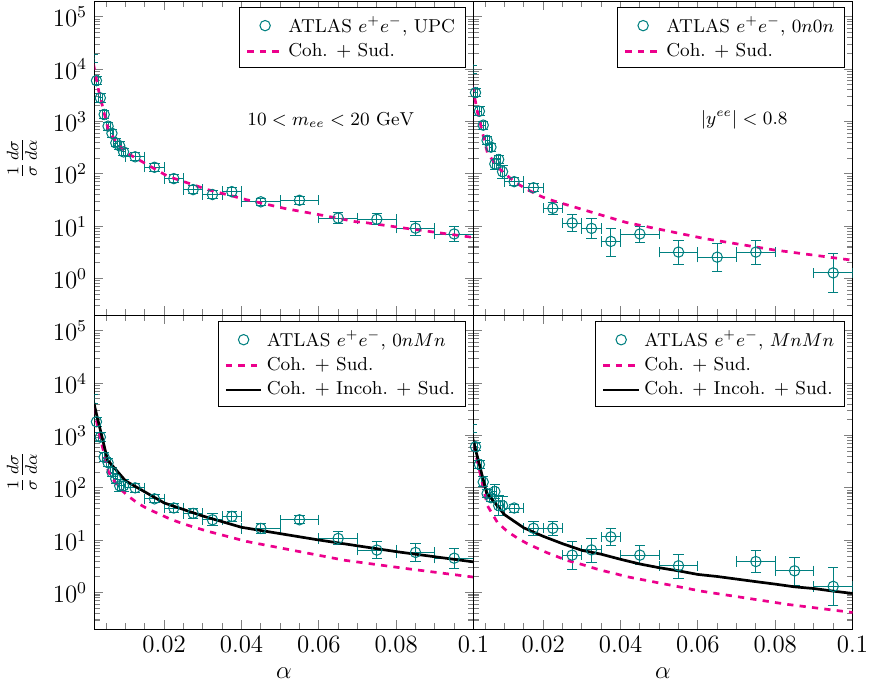}
\caption{Comparing of our results with the ATLAS experimental measurements~\cite{ATLAS:2022srr} of the $e^+e^-$ pair acoplanarity $\alpha$ distribution with neutron emissions in UPCs.}  
\label{fig:atlas_ee_nm08}
\end{figure*} 

When high transverse momentum photons (i.e., incoherent photons) are emitted from the substructure of a nucleon (proton, neutron), it often leads to the breakup of the emitting nucleus. Using the study on events with neutron emissions in the small $\alpha$ region discussed in the last subsection as calibration, we further incorporate the incoherent contributions into our numerical study and explore the production mechanism of the $q_\perp$ broadening effect on the data involving the neutron emission effect in the large $\alpha$ region. As a result, we can demonstrate that the incoherent contributions are indispensable for fully describing the $\alpha$ distribution measurements with neutron emissions.

In the following calculations, since the rest of the observables are not very sensitive to the choice of neutron emission prescription, we will only show numerical results based on  Eq.~(\ref{eq:S}), PAR I, as the prescription to account for the neutron emission effect. 

Firstly, for the events with one or no neutron emission, the outgoing nucleus is almost intact, and the radiated photons are mostly from the nucleus as a whole. Therefore, the radiation of coherent photons should be the dominant contribution in these events. As shown in the above section, the Sudakov resummation, which resums the final-state soft-photon radiations, is the dominant mechanism for the $q_\perp$ broadening in the large $\alpha$ region. In Fig.~\ref{fig:atlas_nm24}, we show the comparison of our calculations involving the Sudakov effect with the ATLAS $0n0n$ data measured in the rapidities regions $|y^{\mu \mu}|<2.4$~\cite{ATLAS:2020epq}. We find that the Sudakov calculation can describe the $0n0n$ data from the ATLAS Collaboration in the large $\alpha$ region. In upper panel of Fig.~\ref{fig:cms_alpha_dis_Sud}, we compare our results incorporating the Sudakov effect with the CMS experimental data~\cite{CMS:2020skx}. For the three scenarios ($0n0n$, $0n1n$, and $1n1n$), we observe that coherent results incorporating the Sudakov effect accurately describe the CMS data in the large $\alpha$ region. Similarly, as indicated in top panel of Fig.~\ref{fig:atlas_mumu_nm08} and Fig.~\ref{fig:atlas_ee_nm08}, the coherent contribution with the Sudakov effect accurately describes both the $0n0n$ and UPC data from ATLAS in dimuon~\cite{ATLAS:2020epq} and dielectron~\cite{ATLAS:2022srr} pair production process. These results serve as the benchmark for our calculation before including the incoherent contributions.

In addition, the neutron emissions indicate that the disassociation of the nucleus when two colliding nuclei get too close to each other in UPCs. In this scenario, the photon might not only be radiated from the nucleus as a whole but also from the protons and quarks inside nucleons. Therefore, the incoherent contribution naturally becomes important for events with more than one neutron emission. Thus, we turn on the incoherent contribution to the photon Wigner distribution. In the lower panel of Fig.~\ref{fig:cms_alpha_dis_Sud}, we present a comparison between our results with and without the incoherent effect and the CMS data across three scenarios ($0nXn$, $1nXn$ and $XnXn$) in dimuon pair production. Similarly, the low panel of Figs.~\ref{fig:atlas_mumu_nm08} and \ref{fig:atlas_ee_nm08} illustrate comparisons of our results with ATLAS data in two scenarios ($0nMn$ and $MnMn$) for dielectron and dimuon pair production. The magenta dashed line represents only the coherent contribution with the Sudakov effect, and the black solid line represents the sum of coherent and incoherent contributions with the Sudakov effect. In the small-$\alpha$ region, both curves agree with each other. It shows that, for the event with neutron emissions,  the coherent contribution, which stands for the radiated photon from the nucleus as a whole, is still the dominant production mechanism of the $q_\perp$ broadening effect in the small $\alpha$ region. In the large $q_\perp$ region, these coherent calculations with the Sudakov effect underestimate the experimental data for tags with neutron emissions, but the sum of the coherent and incoherent contribution with the Sudakov effect provides good descriptions of the data from CMS and ATLAS Collaborations. It means that besides the Sudakov effect, the incoherent contribution, which represents the radiated photon from protons and quarks, become the dominant contribution to the $q_\perp$ broadening effect in the large $\alpha$ regions for events with multiple neutron emissions. 

Let us interpret the numerical results shown in this part. It is evident that the transverse momentum imbalance $q_\perp$ of the lepton pair is small in the small-$\alpha$ region. Thus, the initial heavy nucleus as a unified entity radiates photons with low momenta, and final dileptons do not know the inner structure inside heavy nuclei. In this region, the dominant contribution in the small-$\alpha$ region is from coherent photons. On the contrary, in the large $\alpha$ region where the momentum imbalance of the dilepton pair $q_\perp$ is much larger than $1/R_A$, the coherent contribution to the broadening drops off very rapidly as $q_\perp$ increases. In this case, incoming photons with high transverse momentum are mostly emitted from the sub-structure in heavy nuclei. Consequently, incoherent photons originating from quarks and protons start to play a vital role.

In the end, through the comprehensive and quantitative comparison with the experimental data above, we demonstrate that the incoherent photon contribution is indispensable for describing the data both with and without neutrons emissions simultaneously.

\end{widetext}

\end{document}